\documentclass[%
reprint,
superscriptaddress,
 amsmath,amssymb,
 aps,
 pra,
 twocolumn,
]{revtex4-2}

\usepackage{graphicx}% Include figure files
\usepackage{dcolumn}% Align table columns on decimal point
\usepackage{bm}% bold math
\usepackage[colorlinks]{hyperref}% add hypertext capabilities
\newcommand{\C}{\mathcal{C}}
\newcommand{\N}{\mathcal{N}}
\newcommand{\ra}{\rangle}
\newcommand{\la}{\langle}
\newcommand{\da}{\dagger}

\newcommand{\CNOT}{\text{CNOT}}
\newcommand{\Tof}{\text{Toffoli}}
\newcommand{\hc}{\text{h.c.}}
\newcommand{\disp}{\text{disp}}
\newcommand{\cat}{\text{cat}}
\newcommand{\pc}{\text{pc}}

\newcommand{\opt}{\text{opt}}
\newcommand{\evo}{\text{evo}}
\newcommand{\tot}{\text{tot}}
\usepackage{xcolor}
\usepackage{commath}
\usepackage{makecell}
\usepackage[normalem]{ulem}

\begin{document}

\title{Construction of Bias-preserving Operations for Pair-cat Code}% Force line breaks with \\

\author{Ming Yuan}
\affiliation{Pritzker School of Molecular Engineering, The University of Chicago, Chicago, IL 60637, USA}

\author{Qian Xu}
\affiliation{Pritzker School of Molecular Engineering, The University of Chicago, Chicago, IL 60637, USA}

\author{Liang Jiang}
\affiliation{Pritzker School of Molecular Engineering, The University of Chicago, Chicago, IL 60637, USA}

\date{\today}

\begin{abstract}
Fault-tolerant quantum computation with depolarization error often requires demanding error threshold and resource overhead.
If the operations can maintain high noise bias -- dominated by dephasing error with small bit-flip error -- we can achieve hardware-efficient fault-tolerant quantum computation with a more favorable error threshold. 
Distinct from two-level physical systems, multi-level systems (such as harmonic oscillators) can achieve a desirable set of \emph{bias-preserving} quantum operations while using continuous engineered dissipation or Hamiltonian protection to stabilize to the encoding subspace. 
For example, cat codes stabilized with driven-dissipation or Kerr nonlinearity can possess a set of biased-preserving gates while continuously correcting bosonic dephasing error. However, cat codes are not compatible with continuous quantum error correction against excitation loss error, because it is challenging to continuously monitor the parity to correct photon loss errors.
In this work, we generalize the bias-preserving operations to pair-cat codes, which can be regarded as a multimode generalization of cat codes, to be compatible with continuous quantum error correction against \emph{both} bosonic loss and dephasing errors.
Our results open the door towards hardware-efficient robust quantum information processing with \emph{both} bias-preserving operations and continuous quantum error correction simultaneously correcting bosonic loss and dephasing errors.
\end{abstract}

\maketitle

\section{Introduction}

% Fault-tolerant quantum computation is one of the major pursuits in quantum information community. In general, the error rate for operations on physical qubits should be lower than a small threshold to reach this goal~\cite{Shor1996}. However, if one type of error on physical qubit level is dominant during the whole information processing scheme, we can just put most of our effort protecting against this error. As a result, the error threshold can be increased~\cite{Aliferis2008, Ataides2021} and the resource overhead required can be reduced~\cite{}. So, seeking biased errors in physical systems and preserving the error bias for all operations are highly desirable to make this merit come true.

Quantum information is powerful but fragile due to the presence of noise and imperfections. Quantum error correction can actively correct physical errors and protect the encoded quantum information, by introducing redundancy in physical systems. Fault-tolerant design also ensures that errors during the quantum error correction will not compromise the encoded quantum information, which enables us to accomplish quantum tasks as accurate as possible if the error probability of each gate operation on physical qubits is below certain threshold~\cite{Shor1996,Nielsen2010}.
For generic depolarization errors, however, fault-tolerant quantum computation often requires demanding error threshold and resource overhead, which poses a major challenge with the current technology.
%However, in typical situations the low error threshold and the huge amount of resource overhead are main obstacles towards fault-tolerance. 
% To mitigate those problems, for example, people have developed a lot of encoding method to increase the corresponding error threshold~\cite{}, and made use of ``flag-qubits" to reduce overhead~\cite{}. 

One promising approach to overcome this challenge is to design quantum error correction schemes specific for realistic errors in physical devices.
For example, when physical systems have a biased-noise structure -- one type of error is stronger than all other types of errors~\cite{Aliferis2009} -- we can design efficient quantum error correction schemes to improve error threshold~\cite{Aliferis2008,Li2019,Tuckett2020,Bonilla_ataides2021} and reduce resource overhead~\cite{Xu2022}. Hence, seeking biased-noise structure and preserving the error bias during operations on the physical qubits are highly desirable to make these merits come true.
In practice, however, it is non-trivial to preserve the biased-noise structure for all quantum operations. For example, phase-flip error can be transformed into bit-flip error and vice versa after a Hadamard gate. Moreover, phase-flip error bias cannot be preserved during the execution of a CNOT gate for physical qubits encoded in two-level systems~\cite{Guillaud2019}.

Distinct from two-level physical systems, multi-level systems (such as harmonic oscillators) can encode quantum information with desirable biased-noise structure and bias-preserving quantum operations. For example, we can use harmonic oscillators with cat codes, which encode quantum bit of information using a subspace spanned by two separated coherent states $|\pm \alpha\ra$~\cite{Mirrahimi2014}. 
With specific choice of computational basis of the cat code, the bit-flip error can be exponentially suppressed by the averaged photon number compared with the phase-flip error, which naturally provides the biased-noise structure~\cite{Guillaud2019,Puri2020}. The cat qubits can be stabilized in both driven-dissipative systems~\cite{Mirrahimi2014} and Kerr-nonlinear resonators with 2-photon driving~\cite{Puri2017}. Both of the stabilization protocols with the biased-noise structure have been experimentally demonstrated~\cite{Leghtas2015,Lescanne2020,Grimm2020}. A set of operations which includes CNOT and Toffoli gates for cat qubits with bias-preserving properties has been proposed in both platforms~\cite{Guillaud2019,Puri2020}. Recently, new method to keep noise bias in Kerr cat qubits suppressing heating-induced leakage~\cite{Putterman2022} and new approaches to realize fast and bias-preserving gates in cat code~\cite{Xu_engineering_2022, Gautier2022} have also been proposed. Further, cat qubits can be concatenated into repetition code level, on which a universal gate set for quantum computation can be constructed by using fundamental bias-preserving operations on physical qubits. Concatenation of cat qubits with different types of surface codes has also been investigated under practical consideration~\cite{Chamberland2022,Darmawan2021}. In addition, multicomponent cat codes encoded in a single mode can also be used to protect against photon loss errors~\cite{Mirrahimi2014}. However, the corresponding quantum error correction strategy for all the protocols we mentioned above to suppress the effect from photon losses rely on measuring parity $(-1)^{\hat a^\da \hat a}$, which is hard to be implemented continuously. As a result, extra overhead might be required for those measurements in the middle of the circuits and the following feedback control, which can lead to extra errors and delays.

In our work, we focus on another type of bosonic codes named pair-cat code, which is an important generalization of cat code into multimode bosonic systems~\cite{Albert2019}. For pair-cat code, any photon loss error happening in either mode can be detected by monitoring the photon number difference between the two modes and we can correct them correspondingly, which enables us to perform continuous error correction against photon loss errors. Different from the parity, the photon number difference is much easier to monitor continuously while keeping the stabilization on. Moreover, we need less averaged photon number per mode to get at least the same protection as in the cat code. With all the merits of the pair-cat code, it is natural to ask whether pair-cat code has similar biased-noise structure and whether we can generalize the methods used to construct bias-preserving operations for cat code~\cite{Guillaud2019,Puri2020} into the pair-cat case while keeping the merit of continuous error correction during operations. 

%As a generalization of the previous works on bias-preserving operations for cat codes~\cite{Guillaud2019,Puri2020}, 
%while introducing continuous syndrome monitoring during gate operations, 
In this work, we successfully construct a set of bias-preserving operations for pair-cat codes (including CNOT and Toffoli, sufficient for universal computation in repetition code level), which can be compatible with continuous quantum error correction of both bosonic loss and dephasing errors. 
%This result also suggests people to further look into bosonic codes encoded in multimode systems in order to achieve more favorable error correction properties. 
The paper is organized as follows. In Sec.~\ref{Sec2}, we will go over the basic encoding scheme of the pair-cat code. In Sec.~\ref{Sec3}, we investigate the construction of bias-preserving operation set in both driven-dissipative systems and Hamiltonian systems. We summarize our results in Sec.~\ref{Sec4}. In the Appendices, we summarize some useful properties of pair-cat code itself, including its stabilization, error correction strategy, and optimal error probabilities during the bias-preserving operations we design in the main text.

\section{Pair-cat Code Stabilization}\label{Sec2}

The pair-cat code itself with stabilization in the driven-dissipative systems has been proposed in~\cite{Albert2019}. Here we first summarize basic properties of the code, and then introduce the Hamiltonian protection scheme as a direct generalization from of the cat code.

We first mention the encoding of the cat code for further comparison. By focusing on the subspace spanned by two coherent states $\{|\alpha\ra, |-\alpha\ra\}$, we introduce the states $|\mathcal{C}^\pm_\alpha\ra$ with fixed even or odd photon number parity, where
\begin{equation}
    |\C^\pm_\alpha\ra := \N_\pm(|\alpha\ra \pm |-\alpha\ra).
\end{equation}
Here $\N_\pm=\frac{1}{\sqrt{2(1\pm e^{-2|\alpha|^2})}}$ is the normalization factor. By encoding $|\C^\pm_\alpha\ra$ as the eigenstates of $X$ operator of the cat qubit with eigenvalue $\pm 1$, we can see that in the large $|\alpha|$ limit, $|\pm \alpha\ra$ states serve as the logical $|0\ra$ and $|1\ra$ of the code. Since physical relevant errors, like photon loss, gain and dephasing noise only act locally in the phase space, which make it hard to flip $|\alpha\ra$ to $|-\alpha\ra$ and vice versa, the bit-flip error is naturally suppressed with the choice of our encoding. In fact, it is exponentially suppressed as the increase of average photon number in the resonator compared with phase-flip error.

Then, we consider a system with two bosonic modes and denote them as mode $\hat a$ and $\hat b$. We introduce the pair-coherent state $|\gamma_\Delta\ra$~\cite{Agarwal1988}, which serves as the basic components in pair-cat code. It is defined as the projection of the identical coherent state in two modes $|\gamma,\gamma\ra := |\gamma\ra\otimes|\gamma\ra$ into a subspace with fixed photon number difference between these modes. Specifically, we have
\begin{equation}
    |\gamma_\Delta\ra = \frac{\hat P_\Delta|\gamma,\gamma\ra}{\sqrt{\N_\Delta}},
\end{equation}
where $\N_\Delta = e^{-2|\gamma|^2} I_\Delta(2|\gamma|^2)$ is the normalization factor and $I_\Delta(x)$ is the modified Bessel function of the first kind. $\hat P_\Delta$ is the projection operator which projects states into a subspace with fixed photon number difference $\hat \Delta := \hat n_b - \hat n_a = \Delta$, which means,
\begin{equation}
    \hat P_\Delta := \sum_{n=0}^{+\infty} |n, n+\Delta\ra\la n, n+\Delta| \qquad (\Delta \geq 0).
\end{equation}
The $\Delta < 0$ case can always be defined similarly by performing a SWAP operation between the two modes. From then on for simplicity we assume $\Delta \geq 0$ by default in the following discussions if there is no further comment.

Two merits need to be highlighted for the $|\gamma_\Delta\ra$ state: first, by analogy with the cat code design where $(\hat a^{2} - \alpha^2)|\pm \alpha\ra = 0$, here we have
\begin{equation}
    (\hat a^2 \hat b^2 - \gamma^4)|\gamma_\Delta\ra = (\hat a^2 \hat b^2 - \gamma^4)|(i\gamma)_\Delta\ra = 0.
\end{equation}
Therefore, $\hat a^2\hat b^2 - \gamma^4$ can dissipatively stabilize the pair-cat code space. We note that the pair-cat has a unique advantage over the cat code, which is for any number of photon loss in either mode, it can only change the pair-coherent state into another subspace with different $\Delta$. Notice that
\begin{equation}
    \hat a\hat P_\Delta = \hat P_{\Delta + 1} \hat a, \quad \hat b\hat P_\Delta = \hat P_{\Delta - 1} \hat b,
\end{equation}
we have
\begin{equation}
\begin{split}
    &\hat a^k |\gamma_\Delta\ra = \gamma^k \sqrt{\frac{\N_{\Delta+k}}{\N_{\Delta}}} |\gamma_{(\Delta+k)}\ra, \\
    &\hat b^l |\gamma_\Delta\ra = \gamma^l \sqrt{\frac{\N_{\Delta-l}}{\N_{\Delta}}}|\gamma_{(\Delta-l)}\ra.
\end{split}
\end{equation}
As a result, this type of error syndrome can be easily monitored by measuring $\hat \Delta$, and then we could design strategies to correct it correspondingly. However, this method does not work if the system suffers from loss error $\hat a \hat b$ since $\Delta$ does not change after $\hat a\hat b$ acting on the state. Later we can see that this will give us an uncorrectable error with our encoding method.

To encode the qubit with the pair-coherent states, we use a generalized ``parity" projection operator $\hat Q^{(\Delta)}_\pm$ within each $\Delta$-fixed subspace. Before giving the definition of $\hat Q^{(\Delta)}_\pm$, we first introduce the projection operator $\hat P^a_\pm$ to $\hat a$ mode with fixed parity as
\begin{equation*}
    \hat P^a_\pm := \frac{\hat I \pm (-1)^{\hat n_a}}{2}.
\end{equation*}
Then $\hat Q^{(\Delta)}_\pm$ can be defined as
\begin{equation}
    \hat Q^{(\Delta)}_\pm := \hat P^a_\pm \hat P_\Delta \qquad (\Delta\geq 0).
\end{equation}
As we pointed out that $\Delta < 0$ case can always be defined by performing a SWAP between two modes, we should use the parity in $\hat b$ mode to define $\hat Q^{(\Delta < 0)}_\pm := \hat P^b_\pm \hat P_\Delta$.

\begin{figure}[t]
    \centering
    \includegraphics{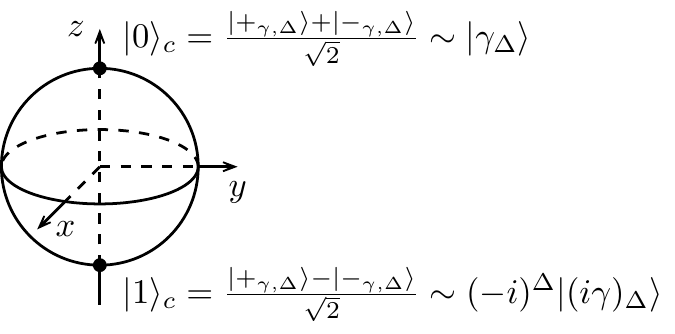}
    \caption{Bloch sphere representation of the pair-cat encoding.}
    \label{fig:Bloch}
\end{figure}

Finally, we define our code states as
\begin{equation}\label{codestates}
    |\pm_{\gamma,\Delta}\ra := \frac{\hat Q^{(\Delta)}_\pm|\gamma, \gamma\ra}{\sqrt{\N_{\pm, \Delta}}} = \frac{|\gamma_\Delta\ra \pm (-i)^\Delta|(i\gamma)_\Delta\ra}{2\sqrt{\N_{\pm,\Delta}/\N_\Delta}},
\end{equation}
where $\N_{\pm, \Delta} = e^{-2|\gamma|^2}[I_\Delta(2|\gamma|^2) \pm J_\Delta(2|\gamma|^2)]/2$ is also a normalization factor, and $J_\Delta(x)$ is the Bessel function of the first kind. We adopt the convention that the above states are eigenstates of the logical $X$ operator, specifically,
\begin{equation}\label{pmc_basis}
    |+\ra_c := |+_{\gamma, \Delta}\ra, \quad |-\ra_c := |-_{\gamma, \Delta}\ra.
\end{equation}
Note that here we use a different choice of basis compared with Ref.~\cite{Albert2019}, so that the phase-flip error is the dominant type of error in our choice of basis in order to be consistent with the existing literature.

In the large $|\gamma|$ limit, like the cat code, we also have $|\la (i\gamma)_\Delta|\gamma_\Delta\ra|^2 \sim O(e^{-4|\gamma|^2})$ , which means these two states are asymptotically orthogonal. As a result,
\begin{equation}\label{asympm}
    |\pm_{\gamma,\Delta}\ra \sim \frac{|\gamma_\Delta\ra \pm (-i)^\Delta|(i\gamma)_\Delta\ra}{\sqrt{2}} \qquad (|\gamma| \to +\infty).
\end{equation}
Further, the states along $Z$ basis are
\begin{equation}\label{asym01}
    |0\ra_c \sim |\gamma_\Delta\ra, \quad |1\ra_c \sim (-i)^\Delta|(i\gamma)_\Delta\ra \qquad (|\gamma| \to +\infty).
\end{equation}

On the other hand, in the $|\gamma| \to 0$ limit, we have
\begin{equation}
\left\{
\begin{aligned}
    &|+_{\gamma,\Delta}\ra \sim (\gamma/|\gamma|)^\Delta |n_a = 0, n_b = \Delta\ra \\
    &|-_{\gamma,\Delta}\ra \sim (\gamma/|\gamma|)^{\Delta + 2} |n_a = 1, n_b = \Delta + 1\ra
    \end{aligned}
\right.
\quad (|\gamma| \to 0),
\end{equation}
% (\gamma/|\gamma|)^\Delta
% \left(\frac{\gamma}{|\gamma|}\right)^\Delta
which, as indicated later, provides us one way to prepare code states in a bias-preserving way by adiabatically turning on control parameters.

As mentioned before, the lowest-order uncorrectable loss error is $\hat a \hat b$. Notice that this error does not cause the code states to go out of the code subspace. We denote $r_\Delta = \sqrt{\N_{-, \Delta}/\N_{+, \Delta}}$. In the large $|\gamma|$ limit, $r_\Delta \simeq 1 - 2e^{-2|\gamma|^2} \cos(\phi)$ where $\phi = 2|\gamma|^2 - \frac{2\Delta + 1}{4}\pi$. The projection operator on the code space can be denoted as $\hat P_c := |+\ra_c\la +| + |-\ra_c\la -|$. In the large $|\gamma|$ limit, we have
\begin{equation}\label{eq:abloss}
\begin{split}
    \hat P_c \hat a \hat b \hat P_c &= \gamma^2 \frac{r_\Delta + r^{-1}_\Delta}{2} \hat Z_c + i\gamma^2 \frac{r_\Delta - r^{-1}_\Delta}{2} \hat Y_c\\
    & \sim \gamma^2 \hat Z_c - 2i \gamma^2 e^{-2|\gamma|^2}\cos(\phi)\hat Y_c.
\end{split}
\end{equation}
We can see that the $Z$ error is the dominant one while the other errors are exponentially suppressed for large $|\gamma|$.

On the other hand, the error induced by bosonic dephasing term can always be exponentially suppressed in the large $|\gamma|$ limit. For example,
\begin{equation}\label{eq:dephasing}
    \hat P_c \hat a^\da \hat a \hat P_c  \sim \gamma^2 \hat I_c +O(\gamma^2 e^{-2|\gamma|^2})\hat X_c.
\end{equation}
Therefore, in our work we will only focus on the effects induced by photon losses and leave the bosonic dephasing error aside.

The pair-cat code can be stabilized in a driven-dissipative system with the jump operator $\hat F = \hat a^2 \hat b^2 - \gamma^4$. The corresponding dynamical equation of motion is
\begin{equation}
    \frac{\dif \rho}{\dif t} = \kappa\mathcal{D}[\hat F]\rho = \kappa(\hat F \rho \hat F^\da - \frac{1}{2}\{\hat F^\da \hat F, \rho\}),
\end{equation}
where $\{\bullet\}$ denotes the anti-commutator.
Note that both the photon number difference and the parity are preserved during this evolution because they commute with $\hat F$. Since $\hat F |\gamma_\Delta\ra = \hat F |(i\gamma)_\Delta\ra = 0$, our code space lies in the decoherence-free subspace of the system. The effective dissipative gap introduced in Ref.~\cite{Albert2017} inversely relates to the leakage rate out of the steady state subspace under perturbations. In our case it has exactly the same properties as the energy gap in the Hamiltonian protection scheme that will be introduced later, and as shown in App.~\ref{AppStru}, it scales as $\Delta_\text{edg} \sim O(|\gamma|^6)$. The property of autonomous error correction of pair-cat code against photon losses is also discussed in~\cite{Albert2019}.

In this work, we suggest that the pair-cat code can also be stabilized by the following Hamiltonian:
\begin{equation}\label{Hamil}
    \hat H = -K(\hat a^{\da 2}\hat b^{\da 2} - \gamma^{*4})(\hat a^2 \hat b^2 - \gamma^4).
\end{equation}
It is easy to see that both $|+_{\gamma, \Delta}\ra$ and $|-_{\gamma, \Delta}\ra$ are the most-excited states (suppose $K > 0$) of this Hamiltonian. Since $[\hat \Delta, \hat H] = [(-1)^{\hat n_a}, \hat H] = [\hat \Delta, (-1)^{\hat n_a}] = 0$, this Hamiltonian can be divided into different parts that act on different subspaces with fixed photon-number difference between two modes and fixed parity:
\begin{equation}\label{eq:Hsep}
    \hat H = \sum_{\mu, \Delta} \hat H_{\mu, \Delta} = \sum_{\mu, \Delta} \hat Q^{(\Delta)}_\mu \hat H \hat Q^{(\Delta)}_\mu.
\end{equation}
The energy gap between the code subspace and first-less-excited states is $\sim 8K|\gamma|^6$ in the large $|\gamma|$ limit when $\Delta$ is a finite number. A more detailed analysis of this Hamiltonian is performed in App.~\ref{AppStru}. 

We also numerically investigated the possibility to find lower order Hamiltonian which has both $\gamma$-dependent protection of the code subspace and preserves the photon number difference. Unfortunately, there is no lower order Hamiltonian that fulfills those criteria. Details can be found in App.~\ref{AppLow}.

\section{Construction of Bias-preserving Gates}\label{Sec3}

The set of bias-preserving operations on cat qubit that does not convert the major errors into the minor errors has been proposed in Ref.~\cite{Guillaud2019,Puri2020}. For single-qubit operations, it contains code state preparation, measurement in $X$ basis, single qubit $X$ operation and rotation along $Z$ axis for arbitrary angle $Z(\theta) := \exp(i\theta \hat Z/2)$. For multi-qubit gates, CNOT and Toffoli gate can be also performed in a bias-preserving manner, which is not possible for physical qubits in two-level systems. Besides, bias-preserving $ZZ(\theta) := \exp(i\theta \hat Z_1 \hat Z_2 / 2)$ gate is also achievable. We denote $\mathcal{S}$ as the set of fundamental bias-preserving operations of the cat code:
$\mathcal{S} = \{\mathcal{P}_{|\pm\ra_c}, \mathcal{M}_X, X, Z(\theta), ZZ(\theta), \CNOT, \Tof \}$. Further, it can also be shown that a universal gate set for fault-tolerant quantum computation can be constructed in the repetition code level by using those bias-preserving operations acting on physical cat qubits. 

In this work, we will show that these operations in $\mathcal{S}$ can also be constructed with the pair-cat code in both driven-dissipative protection and Hamiltonian protection schemes, and reveal the similarities between cat code and pair-cat code structures. The construction of logical gate set on the concatenated code level based on fundamental bias-preserving operations on physical qubits is independent of what the specific type of physical qubits we use, which means the results developed using cat code can be adapted to the pair-cat situation.

The biased error in pair-cat code comes from the large distance between $|\gamma_\Delta\ra$ and $|(i\gamma)_\Delta\ra$ in the generalized phase space (or ``$\gamma$-plane", see~\cite{Albert2019}) and the locality of the physical errors. Therefore, to preserve error bias during the gate operation, $|\gamma|$ should always be kept large.

\begin{table*}
\caption{A summary of fundamental bias-preserving operations in cat code and pair-cat code.}
\begin{ruledtabular}
\begin{tabular}{ccccc}
 &\multicolumn{2}{c}{Cat code}&\multicolumn{2}{c}{Pair-cat code}\\ \hline
 & Driven-dissipative scheme~\cite{Guillaud2019} & Hamiltonian scheme~\cite{Puri2020} & Driven-dissipative scheme & Hamiltonian scheme\\ \hline
 Stabilization & $\hat F_\cat = \hat a^2 - \alpha^2 $ & $\hat H = -K\hat F_\cat^\da \hat F_\cat$ & $\hat F_\pc = \hat a^2 \hat b^2 - \gamma^4 $ & $\hat H = -K\hat F_\pc^\da \hat F_\pc$ \\ \hline
 \makecell{Uncorrectable \\ loss error} & \multicolumn{2}{c}{$\hat a$} & \multicolumn{2}{c}{$\hat a \hat b$} \\ \hline
 $\mathcal{P}_{|+\ra_c}$& \makecell{Start at $|0\ra$ \\ Evolve to steady state} & \makecell{Start at $|0\ra$ \\ $\alpha(t): 0 \to \alpha$} & \makecell{Start at $|0, \Delta\ra$ \\ Evolve to steady state} & \makecell{Start at $|0, \Delta\ra$ \\ $\gamma(t): 0 \to \gamma$} \\ \hline
 $\mathcal{M}_X$ & \multicolumn{2}{c}{\makecell{Need an ancilla: $\hat H_{\text{disp}} = -\chi |e\ra\la e|\hat a^\da \hat a$; \\turn off the protection}} & \multicolumn{2}{c}{\makecell{Need two ancilla: $\hat H_{\text{disp}} = -\chi (|e\ra_1\la e|\hat a^\da \hat a + |e\ra_2\la e|\hat b^\da \hat b)$;\\
 may also need $\hat \Delta$ measurement; turn off the protection}} \\ \hline
 $Z(\theta)$ & \multicolumn{2}{c}{$\hat H_Z = \epsilon_Z (\hat a e^{-i\varphi} + \hat a^\da e^{i\varphi})$} & \multicolumn{2}{c}{$\hat H_Z = \epsilon_Z (\hat a \hat b e^{-i\varphi} + \hat a^\da \hat b^\da e^{i\varphi})$} \\
 $ZZ(\theta)$ & \multicolumn{2}{c}{$\hat H_{ZZ} = \epsilon_{ZZ} (\hat a_1 \hat a_2^\da + \hat a_1^\da \hat a_2)$} & \multicolumn{2}{c}{$\hat H_{ZZ} = \epsilon_{ZZ} (\hat a_1 \hat b_1 \hat a_2^\da \hat b_2^\da + \hat a_1^\da \hat b_1^\da \hat a_2 \hat b_2)$} \\ \hline
 $X$ & \multicolumn{2}{c}{\makecell{$\alpha(t) = \alpha e^{i\pi\frac{t}{T}}$ \\ together with $\hat H_{X, \text{rot}} = -\frac{\pi}{T}\hat n$}}  & \multicolumn{2}{c}{\makecell{$\gamma(t) = \gamma e^{i\frac{\pi}{2}\frac{t}{T}}$ \\ together with $\hat H_{X, \text{rot}} = -\frac{\pi}{2T}(\hat n_a + \hat n_b)$}}\\ \hline
 CNOT& \makecell{$\hat F_{1, 2}$ in Eq.~\eqref{FCNOTcat},\\ with $\hat H_{\CNOT, \text{rot}}$ in Eq.~\eqref{HCNOTcatrot} } & \makecell{Same as Eq.~\eqref{HCNOTpc} with \\ operators for cat code} & \makecell{$\hat F_{1, 2}$ in Eq.~\eqref{FCNOTpc},\\ with $\hat H_{\CNOT, \text{rot}}$ in Eq.~\eqref{HCNOTpcrot}; \\ real-time $\hat \Delta_2$ monitoring } & \makecell{$\hat H_\CNOT$ in Eq.~\eqref{HCNOTpc}; \\ real-time $\hat \Delta_2$ monitoring}  \\ \hline
 Toffoli & \makecell{$\hat F_{1, 2, 3}$ in Eq.~\eqref{FTofcat},\\ with $\hat H_{\text{Tof}, \text{rot}}$ in Eq.~\eqref{HTofcatrot} } & \makecell{Same as Eq.~\eqref{HTofpc} with \\ operators for cat code}  &\makecell{$\hat F_{1, 2, 3}$ in Eq.~\eqref{FTofpc},\\ with $\hat H_{\text{Tof}, \text{rot}}$ in Eq.~\eqref{HTofpcrot}; \\ real-time $\hat \Delta_3$ monitoring } & \makecell{$\hat H_\text{Tof}$ in Eq.~\eqref{HTofpc}; \\ real-time $\hat \Delta_3$ monitoring}\\
\end{tabular}
\end{ruledtabular}
\end{table*}

\subsection{Dissipation Engineering Scheme}\label{SubSec3A}

In this part, we will show the way to construct bias-preserving operations in $\mathcal{S}$ with driven-dissipative stabilization. We will see that how the continuous syndrome ($\hat \Delta$) monitoring can help to reduce errors caused by photon loss. We also derive the scaling properties of the error probability during gate operations where optimal gate time is chosen in App.~\ref{AppError} and summarize the results in TABLE.~\ref{TableErr}.

\textit{Preparation of $|\pm\ra_c$ states}. In cat code protection with jump operator $\hat F_{\text{cat}} = \hat a^2 - \alpha^2$, the preparation of $|\mathcal{C}^+_\alpha\ra$ state can be done by initializing the system at the vacuum state $|0\ra$, and then just let the system evolve under the Lindblad equation to reach the steady state, which will be the exact code state we want~\cite{Mirrahimi2014, Guillaud2019}. It is because that, the steady states $\rho_\infty$ of this evolution is a linear superposition of $\{|\mathcal{C}^+_\alpha\ra\la\mathcal{C}^+_\alpha|, |\mathcal{C}^+_\alpha\ra\la\mathcal{C}^-_\alpha|, |\mathcal{C}^-_\alpha\ra\la\mathcal{C}^+_\alpha|, |\mathcal{C}^-_\alpha\ra\la\mathcal{C}^-_\alpha| \}$. And, since the parity is preserved during the evolution, the only result will only be $\rho_\infty = |\mathcal{C}^+_\alpha\ra\la\mathcal{C}^+_\alpha|$ if the initial state is $|0\ra\la 0|$~\cite{Albert2014}. To prepare $|C^-_\alpha\ra$ state, we can either start with Fock $|1\ra\la 1|$ state and let the system evolve, or perform $\hat Z_c$ operation after the preparation of $|\mathcal{C}^+_\alpha\ra$ state.

Similarly, in pair-cat code case with jump operator $\hat F_\pc = \hat a^2 \hat b^2 - \gamma^4$, the space of steady states is spanned by $\{|\mu_{\gamma,\Delta}\ra\la \mu'_{\gamma,\Delta'}| |\mu, \mu' \in \{+, -\}; \Delta, \Delta' \in \mathbb{Z} \}$. Besides, both the parity of the two modes and the photon number difference are conserved. As a result, if we start with $|0,\Delta\ra\la 0,\Delta|$ state and let the system evolve, eventually it will end up at $|+_{\gamma, \Delta}\ra\la +_{\gamma, \Delta}|$ state. To prepare $|-_{\gamma,\Delta}\ra$ state, similarly we can either start with $|1, \Delta + 1\ra$ state and wait for it to reach the steady state, or perform $\hat Z_c$ operation, which we will introduce later, on $|+_{\gamma, \Delta}\ra$ state. 

Photon loss errors may happen during the state preparation and the idling time after that. The probability of a single-photon loss in either mode is $p = \kappa_1 \bar n T$ provided that $p \ll 1$, where $\kappa_1$ is the 1-photon loss rate, $T$ is the total time of the process we consider, and $\bar n$ is the average photon number during the whole process. However, as indicated before, only a single-photon loss does not cause a logical error directly. It can be captured by a $\hat \Delta$ measurement after the process to determine whether and in which mode the photon loss happens.  Then we can apply a recovery operation to correct the error. However, loss errors happen in both modes cannot be identified in this way, which can occur with the probability $p_Z = p^2 \sim O[(\kappa_1 \bar n T)^2]$. This corresponds to the $Z$ error probability of the pair-cat code during state preparation and idling process. In the idling part, we have $\bar n \sim O(|\gamma|^2)$, and if the time of this part dominates we can write $p_Z = p^2 \sim O[(\kappa_1 |\gamma|^2 T)^2]$.

\textit{Measurement in $X$ basis}. In order to distinguish $|+_{\gamma,\Delta}\ra$ state with $|-_{\gamma, \Delta}\ra$ state, we can try to check the parity of either mode of the pair-cat code. This can be done in the same way as the cat code case~\cite{Guillaud2019}. We could couple $\hat a$ mode with an ancilla qubit via dispersive coupling Hamiltonian $\hat H_{\text{disp}} = -\chi |e\ra\la e|\hat a^\da \hat a$. The ancilla qubit is initialized at $|+\ra_q$ state where $|\pm\ra_q = (|g\ra \pm |e\ra)/\sqrt{2}$. After time $T = \frac{\pi}{\chi}$, the unitary evolution operator will be $\hat U = |g\ra\la g|\otimes \hat I + |e\ra\la e|\otimes e^{i\pi \hat n_a}$, and the quantum state will evolve from $|\psi(0)\ra = |+\ra_q \otimes(u_0|+_{\gamma, \Delta}\ra + u_1|-_{\gamma, \Delta}\ra)$ to $|\psi(T)\ra = u_0|+\ra_q \otimes |+_{\gamma, \Delta}\ra + u_1|-\ra_q \otimes|-_{\gamma, \Delta}\ra$. Finally, we measure the ancilla qubit along $X$ basis. If we get $|+\ra_q$ state, it is equivalent to say that we get the $|+_{\gamma, \Delta}\ra$ by performing the $X$ measurement on pair-cat code.

It is worth to mention that, during the qubit-dependent rotation of the cavity modes, we have
\begin{equation}
    e^{i\theta\hat n_a}|\gamma_\Delta\ra = e^{-i\Delta\theta/2}|(\gamma e^{i\theta/2})_\Delta\ra,
\end{equation}
% which implies that the magnitude of $|\gamma e^{i\theta/2}| = |\gamma|$ is kept as a constant. which provides exponential suppression of bit-flip error of pair-cat code in large $|\gamma|$ limit all the time during measurement.
which means the driven-dissipative stabilization should be turned off during this evolution. 
%\ming{[MY: Change to the following to resolve ``this" \& ``that" issue]} 
However, as indicated in the cat code case~\cite{Guillaud2019}, turning off the stabilization is not a problem because the only information we need from the measurement is the parity of the state instead of the amplitude $\gamma$. Moreover, Since the dissipator $\hat F$ commutes with the parity, it does not provide any protection against parity change. As a result, it does not matter whether the dissipative stabilization is on or off during the measurement process.

% \color{red}
% However, as indicated in the cat code case~\cite{Guillaud2019}, this is not a problem since the only information we need is the parity of the state instead of the amplitude $\gamma$, and the dissipator $\hat F$ commutes with the parity, which means it does not provide protection against parity change.

% \textcolor{green}{In fact, we can also use a second ancilla qubit to perform the parity measurement of the $\hat b$ mode together with the $\hat a$ mode. The advantage of using two ancilla is, suppose during the dispersive evolution process some photon loss errors happen in either of the two modes (but not both), then the measurement outcome of the ancilla coupled to that mode cannot be trusted. However, the parity of the other mode does not change, so the outcome of the parity measurement of that mode can still be used. Since loss error only happen in one mode, we can finally measure $\hat \Delta$ to see which mode suffers from the loss while the other one does not. Notice that $\hat \Delta$ and $\hat X_q$ operators of the two ancilla commute with each other, it does not matter which one is measured first. }(Need to be modified)

Note that a single photon loss might change the outcome of the parity measurement. To suppress the loss-induced measurement error to higher order, we can introduce two ancilla and use them to measure both the parity of $\hat a$ and $\hat b$ mode together. If the outcomes agree with the $\Delta$ we fixed for the code space, we can trust the outcomes. Otherwise, we need to 
perform a measurement of the photon number difference between the two modes immediately after the parity measurement, to check which mode suffers from the photon loss and use the parity of another mode to indicate the generalized parity of the pair-cat code state.
% [LJ: What if the photon loss happens during the number difference measurement? How about we perform parity measurement for both cavities and compare the outcomes? If disagree, we can perform number difference measurement to identify the mode that suffers from the loss error, and resolve the parity discrepency. This will guarantee that the loss-induced parity measurement is suppressed to the higher order.]
% \color{black}
However, if both modes suffer from single-photon loss during parity measurement, there is some chance that the parity outcomes are consistent but wrong, or they are inconsistent but cannot be resolved since $\hat \Delta$ measurement suggests no loss happened. As a result, the error probability during the measurement process can be suppressed from $O(\kappa_1 |\gamma|^2/\chi)$ to $O[(\kappa_1 |\gamma|^2/\chi)^2]$ by using the protocol we mentioned here.

% It is also equivalent to have one parity measurement of either mode and another measurement of photon number difference between two modes.

\textit{$Z(\theta)$ and $ZZ(\theta)$ gates}. The $Z(\theta)$ and $ZZ(\theta)$ gate in cat code can be performed by using the following Hamiltonian~\cite{Mirrahimi2014}:
\begin{equation}
\begin{split}
    &\hat H_Z = \epsilon_Z (\hat a e^{-i\varphi} + \hat a^\da e^{i\varphi}),\\
    &\hat H_{ZZ} = \epsilon_{ZZ} (\hat a_1 \hat a_2^\da + \hat a_1^\da \hat a_2).
\end{split}
\end{equation}
By projecting those Hamiltonian into the cat code subspace, we can get the $Z$ and $ZZ$ operator which will generate the $Z(\theta)$ and $ZZ(\theta)$ gates. Here $\theta$ can be controlled by the gate time. The validity of this projection can be understood via quantum Zeno effect, that the dissipation term keeps monitoring the system to prevent the state from leaking out of the code subspace, given that $\epsilon_Z, \epsilon_{ZZ} \ll \kappa$.
Since both of the projected Hamiltonian commute with $Z$ error on either cat qubit, these two operations are naturally bias-preserving.

In the pair-cat situation, we can use these Hamiltonian to achieve the two gates~\cite{Albert2019}:
\begin{equation}\label{HZHZZ}
\begin{split}
    &\hat H_Z = \epsilon_Z (\hat a \hat b e^{-i\varphi} + \hat a^\da \hat b^\da e^{i\varphi}),\\
    &\hat H_{ZZ} = \epsilon_{ZZ} (\hat a_1 \hat b_1 \hat a_2^\da \hat b_2^\da + \hat a_1^\da \hat b_1^\da \hat a_2 \hat b_2).
\end{split}
\end{equation}
Again, by projecting into the code space with $\hat P_c = |+\ra_c\la +| + |-\ra_c\la -|$ while working in the large $|\gamma|$ limit, we have
\begin{subequations}\label{PHZHZZ}
\begin{equation}\label{PHz}
\begin{split}
    &\hat P_c \hat H_Z \hat P_c \\
    \sim{}& 2\epsilon_Z\left(\Re[\gamma^2 e^{-i\varphi}] \hat Z_c + 2\Im[\gamma^2 e^{-i\varphi}] e^{-2|\gamma|^2}\cos(\phi) \hat Y_c\right),
    \end{split}
\end{equation}
\begin{equation}
    \begin{split}
    &(\hat P_{1c} \hat P_{2c}) \hat H_{ZZ} (\hat P_{1c} \hat P_{2c})\\
    \sim{}& 2\epsilon_{ZZ}|\gamma|^4 (\hat Z_{1c}\hat Z_{2c} + 4e^{-4|\gamma|^2}\cos^2(\phi)\hat Y_{1c} \hat Y_{2c}).
    \end{split}
\end{equation}
\end{subequations}
Here $\phi = 2|\gamma|^2 - \frac{2\Delta + 1}{4}\pi$ has been introduced before. We can also see that the $\hat Y_c$ and $\hat Y_{1c} \hat Y_{2c}$ terms are exponentially suppressed so that we can use these Hamiltonian to get $Z(\theta)$ and $ZZ(\theta)$ gates. Besides, in Eq.~\eqref{PHz} we can always choose $\varphi$ so that $\gamma^2 e^{-i\varphi} = |\gamma|^2$. The corresponding gate time to reach $\theta$ angle rotation is $t_Z = \frac{\theta}{4|\epsilon_Z||\gamma|^2}$ and $t_{ZZ} = \frac{\theta}{4|\epsilon_{ZZ}||\gamma|^4}$, where to be consistent $\epsilon_Z$ and $\epsilon_{ZZ}$ should be chosen as $\epsilon_Z = -|\epsilon_Z|$ and $\epsilon_{ZZ} = -|\epsilon_{ZZ}|$.

\textit{$X$ gate}. To realize $X$ gate in cat code in a bias-preserving way, one method is to adiabatically change $\alpha(t)$ from $\alpha$ to $-\alpha$ and vice versa, while keeping $|\alpha(t)|$ larger all the time to protect the error bias~\cite{Guillaud2019}. After that, $|\C^+_\alpha\ra$ state will remain as $|\C^+_\alpha\ra$, while $|\C^-_\alpha\ra$ changes to $-|\C^-_\alpha\ra$, which is exactly the outcome of $X$ gate acting on code states.

In pair-cat code case, we can also let $\gamma(t)$ change adiabatically along $\gamma(t) = \gamma e^{i\frac{\pi}{2}\frac{t}{T}}$ from $t=0$ to $T$. In this way, $|\gamma_\Delta\ra$ goes to $|(i\gamma)_\Delta\ra$ while $|(i\gamma)_\Delta\ra$ goes to $|(-\gamma)_\Delta\ra = (-1)^\Delta |\gamma_\Delta\ra$. As a result,
\begin{equation}
    |+_{\gamma, \Delta}\ra \to i^\Delta |+_{\gamma, \Delta}\ra, \quad |-_{\gamma, \Delta}\ra \to i^\Delta (- |-_{\gamma, \Delta}\ra).
\end{equation}
So, equivalently, this is a $e^{i\Delta \pi/2} \hat X_c$ operation, while the global phase does not matter.

In order to implement this design in physical systems, we need to engineer the jump operator as $\hat F = \hat a^2 \hat b^2 - \gamma^4(t)$. We can also add a Hamiltonian $\hat H_{X, \text{rot}} = -\frac{\pi}{2T}(\hat n_a + \hat n_b)$. It can be checked that
\begin{equation}
\begin{split}
    &\exp(-i\hat H_{X, \text{rot}} t)(u_0|\gamma_\Delta\ra + u_1 |(i\gamma)_\Delta\ra)\\
    ={}& u_0|(\gamma e^{i\frac{\pi}{2}\frac{t}{T}})_\Delta\ra + u_1 |(i\gamma e^{i\frac{\pi}{2}\frac{t}{T}})_\Delta\ra,
    \end{split}
\end{equation}
which means that this state is always annihilated by $\hat F = \hat a^2 \hat b^2 - \gamma^4(t)$, so that it is protected by the driven-dissipative stabilization for any time during gate execution. So, with the help of this Hamiltonian, we could relax the requirement of adiabaticity such that $T \to +\infty$ is not needed.

\textit{CNOT gate}. The idea of implementing a CNOT gate is similar to the $X$ gate: since $\CNOT = |0\ra\la 0|\otimes \hat I + |1\ra\la 1|\otimes \hat X$, we adiabatically rotate the target mode conditioned on the control mode being in the $|1\ra$ state. Therefore, in cat code scheme, the jump operators of these two cat qubits was proposed~\cite{Guillaud2019} as
\begin{equation}\label{FCNOTcat}
    \hat F_1 = \hat a^2_1 - \alpha^2, \quad \hat F_2 = \hat a^2_2 - \frac{\alpha(\hat a_1 + \alpha)}{2} + \frac{\alpha e^{2i\pi t/T} (\hat a_1 - \alpha)}{2},
\end{equation}
where, in the large $|\alpha|$ limit, by fixing control qubit in its code space, we have $\hat F_2 \sim |\alpha\ra\la \alpha|\otimes (\hat a_2^2 - \alpha^2) + |-\alpha\ra\la -\alpha|\otimes (\hat a_2^2 - \alpha^2(t))$ where $\alpha(t) = \alpha e^{i\pi t/T}$. So, when $|\alpha|$ is large, if the control qubit is in $|\alpha\ra$ state which is encoded as $|0\ra_c$ asymptotically, the state of the target qubit does not change; on the other hand, if the control qubit is in $|-\alpha\ra$ state, effectively there will be an $X$ operation acting on the target qubit.

Again, like the $X$ gate construction, we can add a Hamiltonian to generate the conditioned rotation of the target qubit to partially compensate the error from non-adiabaticity:
\begin{equation}\label{HCNOTcatrot}
\begin{split}
    \hat H_{\CNOT, \text{rot}} &= \frac{\pi}{2T}\frac{\hat a_1 - \alpha}{2\alpha} \otimes (\hat a_2^\da \hat a_2 - |\alpha|^2) + \hc\\
    & \sim -\frac{\pi}{T} |-\alpha\ra\la -\alpha|\otimes (\hat a_2^\da \hat a_2 - |\alpha|^2).
\end{split}
\end{equation}

To achieve the actual CNOT operation, we need an extra $Z_1(-\pi |\alpha|^2)$ gate acting on the control qubit~\cite{Chamberland2022}. In fact, we can choose $|\alpha|^2$ as an even integer so that This extra action is not needed.

In pair-cat code case, we use the following jump operators to stabilize the code states:
\begin{equation}\label{FCNOTpc}
    \begin{split}
        & \hat F_1 = \hat a_1^2 \hat b_1^2 - \gamma^4,\\
        & \hat F_2 = \hat a_2^2 \hat b_2^2 - \frac{\gamma^2 (\hat a_1\hat b_1 + \gamma^2)}{2} + \frac{\gamma^2 e^{2i\pi t/T}(\hat a_1\hat b_1 - \gamma^2)}{2}.
    \end{split}
\end{equation}
And the Hamiltonian we need for partially compensating the non-adiabatic error is
\begin{equation}\label{HCNOTpcrot}
    \hat H_{\CNOT, \text{rot}} = \frac{\pi}{4T}\frac{\hat a_1\hat b_1 - \gamma^2}{2\gamma^2} \otimes (\hat a_2^\da \hat a_2 + \hat b_2^\da \hat b_2 - 2|\gamma|^2) + \hc.
\end{equation}
However, the extra phase induced during $\gamma$ rotation should be taken into consideration, since our effective gate operator now is $\hat U \propto |0\ra_{1c}\la 0|\otimes \hat I_{2c} + |1\ra_{1c}\la 1|\otimes e^{-i\pi(|\gamma|^2 - \Delta/2)}\hat X_{2c}$. We can always use $Z(\theta)$ gate on the control qubit to correct the induced phase, or choose $\gamma$ such that $(|\gamma|^2 - \Delta/2)$ is an even integer.

Different from cat code where a single-photon loss can cause a phase-flip error on cat qubit, we seek for protocols that for the pair-cat code a single-photon loss during the CNOT gate execution does not cause logical errors. If we do nothing more than what is discussed above, we will not know when a single-photon loss event might happen, which will lead to a $Z$ type of error on the control qubit if a photon loss occurs on target qubit. For example, we assume an $\hat a_2$ error happens at time $t_0$ on the target qubit and see what the code states will become finally. We still consider the large $|\gamma|$ regime where Eq.~\eqref{asym01} is satisfied, and use $|0_\Delta\ra_c$ and $|1_\Delta\ra_c$ to denote the code states defined in specified $\Delta$ subspace. Approximately we have
\begin{equation}\label{qjump}
    \begin{split}
        & |0_\Delta\ra_{1c}|0_\Delta\ra_{2c} \to \gamma  |0_\Delta\ra_{1c}|0_{\Delta+1}\ra_{2c},\\
        & |0_\Delta\ra_{1c}|1_\Delta\ra_{2c} \to (-1)\gamma |0_\Delta\ra_{1c}|1_{\Delta+1}\ra_{2c},\\
        & |1_\Delta\ra_{1c}|0_\Delta\ra_{2c} \to i^{\Delta + 1}\gamma(t_0)  |1_\Delta\ra_{1c}|1_{\Delta+1}\ra_{2c},\\
        & |1_\Delta\ra_{1c}|1_\Delta\ra_{2c} \to (-1)i^{\Delta + 1}\gamma(t_0)  |1_\Delta\ra_{1c}|0_{\Delta+1}\ra_{2c}.
    \end{split}
\end{equation}
Here we just omit some overall factors which are the same for all the final states in the expressions above. After the evolution, the final states should go through a recovery channel by syndrome ($\hat \Delta$) measurement and error correction. We have a more detailed discussion in App.~\ref{AppQEC} on the recovery strategy based on the outcome of the final $\Delta$ we measured. Briefly, the recovery process will map $|0_{\Delta +1}\ra_{2c}$ to $|0_\Delta\ra_{2c}$ and map $|1_{\Delta + 1}\ra_{2c}$ to $(-1)|1_\Delta\ra_{2c}$ for the target qubit.

% \color{red}

% [LJ: Since the phase is $\exp[i\frac{\pi}{2}(\Delta -2|\gamma|^2 )] \times \exp[-i\frac{\pi}{2}(1 - t_0/T)]$, can first term does not depend on $t_0$, can we just remove it so simplify the expression?]
% \color{black}
%\ming{[MY: solve the over-complicated expression issue]} \color{brown}
After the recovery, if the control qubit is in $|1_\Delta\ra_{1c}$, then in addition to the $e^{-i\pi(|\gamma|^2 - \Delta/2)}$ phase that will be achieved in the no error case we mentioned above, there will be an extra $\exp[-i\frac{\pi}{2}(1 - t_0/T)]$ phase on the final states, since $\gamma(t) = \gamma \exp(i\frac{\pi}{2}\frac{t}{T})$. So, if we do not know what $t_0$ is, this induced time-dependent phase cannot be corrected.

Indeed, this CNOT gate is still bias-preserving, since the error induced by single-photon loss is still $Z$ type of error, which is the dominant one. However, it violates one of the proposed merits of pair-cat code that the single-photon loss error in either mode will not cause errors in the code. To solve this issue, one method is to introduce real-time monitoring of photon number difference $\hat \Delta_2$ on the target qubit to keep track of the time when the loss error might happen. It is in principle doable since $\hat \Delta_2$ commutes with all the generators in the CNOT gate design and the code states will not be changed during measurement since they are always eigenstates of $\hat \Delta_2$, regardless of whether they suffer from loss errors or not. If we know the specific time that the single-photon loss happens, we can apply a $Z_1(\theta)$ gate on the control qubit to correct the induced phase. Therefore, the leading uncorrectable error will again be suppressed to higher order, which comes from both the inaccuracy of the phase correction due to the finite time interval of different $\hat \Delta$ measurement, and the situation that both $\hat a$ and $\hat b$ error happen in the same time interval between two $\hat \Delta$ measurement. We have a detailed analysis of those gate errors in App.~\ref{AppError}. It is worth to mention that, in the limit that the time interval of two $\hat \Delta$ measurement can be ignored, due to the large dissipation gap the optimal CNOT gate error probability will decrease as $\gamma$ increases. This is in contrast to the cat code case where the optimal error probability of CNOT is independent of the size of the cat states.

\textit{Toffoli gate}. Since the Toffoli gate is just the Control-CNOT gate, we can extend the strategy introduced in the construction of CNOT gate for the Toffoli case. For the cat code, the jump operators and rotation Hamiltonian have been proposed as~\cite{Guillaud2019}
\begin{subequations}\label{FTofcat}
\begin{equation}
    \hat F_1 = \hat a_1^2 - \alpha^2, \quad \hat F_2 = \hat a_2^2 - \alpha^2,
\end{equation}
\begin{equation}
\begin{split}
    \hat F_3 ={}& \hat a_3^2 - \frac{1}{4}(\hat a_1 + \alpha)(\hat a_2 + \alpha) +\frac{1}{4}(\hat a_1 - \alpha)(\hat a_2 + \alpha)\\
    &+ \frac{1}{4}(\hat a_1  + \alpha)(\hat a_2 - \alpha) - \frac{1}{4} e^{2i\pi\frac{t}{T}}(\hat a_1 - \alpha)(\hat a_2 - \alpha),
\end{split}
\end{equation}
\end{subequations}
with
\begin{equation}\label{HTofcatrot}
\begin{split}
    \hat H_{\text{Tof}, \text{rot}} ={}& -\frac{\pi}{2T}\left(\frac{\hat a_1 - \alpha}{2\alpha}\otimes\frac{\hat a_2^\da - \alpha^*}{2\alpha^*} + \hc\right)\\
    & \otimes (\hat a_3^\dagger \hat a_3 - |\alpha|^2).
\end{split}
\end{equation}

While, in the pair-cat code case, the jump operators can be chosen as
\begin{subequations}\label{FTofpc}
\begin{equation}
    \hat F_1 = \hat a_1^2 \hat b_1^2 - \gamma^4, \quad \hat F_2 = \hat a_2^2 \hat b_2^2 - \gamma^4,
\end{equation}
\begin{equation}\label{FTofpc3}
\begin{split}
    \hat F_3 ={}& \hat a_3^2 \hat b_3^2 - \frac{1}{4}(\hat a_1 \hat b_1 + \gamma^2)(\hat a_2 \hat b_2 + \gamma^2)\\
    &+\frac{1}{4}(\hat a_1 \hat b_1 - \gamma^2)(\hat a_2 \hat b_2 + \gamma^2)\\
    &+ \frac{1}{4}(\hat a_1 \hat b_1 + \gamma^2)(\hat a_2 \hat b_2 - \gamma^2)\\
    &- \frac{1}{4} e^{2i\pi\frac{t}{T}}(\hat a_1 \hat b_1 - \gamma^2)(\hat a_2 \hat b_2 - \gamma^2).
\end{split}
\end{equation}
\end{subequations}
Besides, the Hamiltonian to compensate the non-adiabatic error is
\begin{equation}\label{HTofpcrot}
\begin{split}
    \hat H_{\text{Tof}, \text{rot}} ={}& -\frac{\pi}{4T}\left(\frac{\hat a_1\hat b_1 - \gamma^2}{2\gamma^2}\otimes\frac{\hat a^\da_2\hat b^\da_2 - \gamma^{*2}}{2\gamma^{*2}} + \hc\right)\\
    & \otimes (\hat a_3^\dagger \hat a_3 + \hat b_3^\dagger \hat b_3 - 2|\gamma|^2).
\end{split}
\end{equation}

Some extra work in CNOT gate construction should also be done here. The induced phase during the rotation of target qubit can be corrected by applying both $Z(\theta)$ and $ZZ(\theta)$ gates on the two control qubits, or just use carefully chosen $\gamma$ such that this phase has no effect. Besides, we need real-time monitoring of $\hat \Delta_3$ on the target qubit to correct the error induced by single-photon loss on that qubit.

\begin{table}
\caption{The scaling of optimal total Z type of error probability in bias-preserving gates for cat code and pair-cat code (with real-time $\hat \Delta$ monitoring on each pair-cat qubits). 
% \color{red}
% [LJ: Express the errors for pair-cat code in terms of $\sqrt{\frac{\kappa_1}{\kappa}}\times \sqrt{\frac{\kappa_1 \delta\tau}{\gamma^2}} \times \cdots$, which is easier to see the improvement comes from $\sqrt{\frac{\kappa_1 \delta\tau}{\gamma^2}}\ll 1$]
% \color{black}
}
\begin{ruledtabular}
\begin{tabular}{ccc}\label{TableErr}
 &Cat code~\cite{Chamberland2022} & Pair-cat code \\ \hline
 Z($\theta$) & $O(\frac{1}{\alpha}\sqrt{\frac{\kappa_1}{\kappa}})$ & $O(\frac{\kappa_1}{\gamma^2}\sqrt{\frac{\delta \tau}{\kappa}})$ \\ \hline
 ZZ($\theta$) & $O(\frac{1}{\alpha}\sqrt{\frac{\kappa_1}{\kappa}})$ & $O(\frac{\kappa_1}{\gamma^2}\sqrt{\frac{\delta \tau}{\kappa}})$ \\ \hline
X & $O(\kappa_1\alpha^2 T) \xrightarrow{T\to 0} 0$ & $O(\kappa_1^2 \gamma^4 \delta \tau T) \xrightarrow{T\to 0} 0$ \\ \hline
CNOT & $O(\sqrt{\frac{\kappa_1}{\kappa}})$ & $O(\sqrt{\frac{\kappa_1}{\kappa}} \sqrt{\frac{\kappa_1 \delta\tau}{\gamma^2}}\sqrt{ 1 + C \kappa\kappa_1\gamma^8 (\delta\tau)^2})$ \\ \hline
Toffoli & $O(\sqrt{\frac{\kappa_1}{\kappa}})$ & $O(\sqrt{\frac{\kappa_1}{\kappa}} \sqrt{\frac{\kappa_1 \delta\tau}{\gamma^2}}\sqrt{ 1 + C' \kappa\kappa_1\gamma^8 (\delta\tau)^2})$
\end{tabular}
\end{ruledtabular}
\end{table}

\subsection{Hamiltonian Stabilization Scheme}\label{SubSec3B}

We note that in some way Hamiltonian stabilization scheme is similar as the dissipative stabilization scheme. We have already got a sense of such similarity from the structure of stabilization Hamiltonian $\hat H = -K \hat F^\da \hat F$ where $\hat F = \hat a^2 \hat b^2 - \gamma^4$ is the jump operator we use in the dissipative stabilization scheme. We can make use of such similarities to construct bias-preserving operations in Hamiltonian stabilization scheme.

\textit{Preparation of $|\pm\ra_c$ states}. To prepare $|+\ra_c$ state of pair-cat code, we can use a similar method as the state preparation in Kerr-cat scheme proposed in ~\cite{Puri2020}. Since $|\pm_{\gamma, \Delta}\ra$ are always the most excited eigenstates of the Hamiltonian shown in Eq.~\eqref{Hamil}, and in $|\gamma| \to 0$ limit we have  $|+_{\gamma,\Delta}\ra \sim |0, \Delta\ra$ and $|-_{\gamma,\Delta}\ra \sim |1, \Delta+1\ra$, we can first prepare $|0, \Delta\ra$ or $|1, \Delta+1\ra$ and adiabatically increase $\gamma(t)$ from $0$ to the final $\gamma$ we want. Since both $\hat \Delta$ and parity are conserved, we will reach the corresponding $|\pm_{\gamma, \Delta}\ra$ state finally in the adiabatic limit.

\textit{Measurement in $X$ basis}. This can be done in the same way as proposed in the driven-dissipative scheme since the protection has to be turned off during the measurement process.

\textit{$Z(\theta)$ and $ZZ(\theta)$ gates}. We can still use the same Hamiltonian in Eq.~\eqref{HZHZZ} to generate $Z(\theta)$ and $ZZ(\theta)$ accordingly. It is because that the Hamiltonian in Eq.~\eqref{Hamil} could provide the protection of the code space because of the $O(|\gamma|^6)$ energy gap, and according to Eq.~\eqref{PHZHZZ}, within the code space $\hat H_Z$ and $\hat H_{ZZ}$ serve as the generators of $Z(\theta)$ and $ZZ(\theta)$ gates.

\textit{$X$, CNOT and Toffoli gates}. The ideas for construction of these three bias-preserving  operations are quite similar: they all require conditioned adiabatically changing of stabilization parameter $\gamma(t)$ while keeping $|\gamma(t)|$ large all the time, and use another Hamiltonian to actively change the code states to reduce the error from non-adiabaticity due to the finite evolution time.

% So, for the $X$ gate, with the jump operator $\hat F = \hat a^2 \hat b^2 - \gamma^4(t)$ where $\gamma(t) = \gamma e^{i\frac{\pi}{2}\frac{t}{T}}$ and Hamiltonian $\hat H_{X, \text{rot}} = -\frac{\pi}{2T}(\hat n_a + \hat n_b)$, as introduced when constructing $X$ gate in driven-dissipative stabilization scheme, here we introduce
% \begin{equation}
%     \hat H_X = -K \hat F^\da \hat F + \hat H_{X, \text{rot}},
% \end{equation}
% where the first term provides stabilization of the code space and the second term actively change code states.

So, we can use the following Hamiltonian to implement $X$ gate,
\begin{equation}
    \hat H_X = -K \hat F^\da \hat F + \hat H_{X, \text{rot}},
\end{equation}
where $\hat F = \hat a^2 \hat b^2 - \gamma^4(t)$ with $\gamma(t) = \gamma e^{i\frac{\pi}{2}\frac{t}{T}}$ in the first term provides stabilization of the code space and the second term $\hat H_{X, \text{rot}} = -\frac{\pi}{2T}(\hat n_a + \hat n_b)$ can actively change code states according to $\gamma = \gamma(t)$ to compensate the error induced by non-adiabaticity.

For the CNOT gate, we can use the following $\hat H_\CNOT$:
\begin{equation}\label{HCNOTpc}
    \hat H_\CNOT = -K(\hat F_1^\da \hat F_1 + \hat F_2^\da \hat F_2) + \hat H_{\CNOT, \text{rot}},
\end{equation}
where $\hat F_1$ and $\hat F_2$ is defined in Eq.~\eqref{FCNOTpc} to provide stabilization and Hamiltonian $\hat H_{\CNOT, \text{rot}}$ is defined in Eq.~\eqref{HCNOTpcrot} to provide mitigation of non-adiabatic error.

The Toffoli gate can be constructed with $\hat H_\text{Tof}$:
\begin{equation}\label{HTofpc}
    \hat H_\text{Tof} = -K \sum_{j=1}^3 \hat F_j^\da \hat F_j + \hat H_{\text{Tof}, \text{rot}}, 
\end{equation}
where $\hat F_j$ is defined in Eq.~\eqref{FTofpc} and Hamiltonian $\hat H_{\text{Tof}, \text{rot}}$ is defined in Eq.~\eqref{HTofpcrot}.

Same as the driven-dissipative case, the real-time $\hat \Delta$ monitoring on target qubits and phase correction on control qubits in both CNOT and Toffoli gates are also needed here.

\section{Discussion and Conclusion}\label{Sec4}

It is possible to generalize the pair-cat encoding protocol into a multimode multicomponent case in order to correct more photon loss and gain errors~\cite{Albert2019}. In general, we could stabilize a $d$ level qudit in $M$ modes with jump operator $\hat F = (\hat a^d)^{\otimes M} - \gamma^{dM}$, and syndromes can be monitored by measuring all the photon number differences between neighboring modes. In this way, any amount of photon loss happening in arbitrary $M-1$ modes can be distinguished, or if $M \geq 3$ then any amount of photon loss or gain happening in $\frac{M-1}{2}$ modes corresponds to a unique syndrome. But there will be a logical error on the qudit if all of the modes suffer from a photon loss together, provided that there is no further encoding on the logical qudit within the $d$ level subspace.

For the multimode pair-cat qubit case ($d = 2$), it is straightforward to achieve the bias-preserving operations from the generalization of 2-mode pair-cat code, just as the generalization from cat code to 2-mode pair-cat. It will be tricky to talk about bias-preserving in cat or pair-cat qudits and their future concatenations, since different single-qudit error may correspond to different number of photon loss or gain which can happen with different probability. But, still the continuous monitoring of syndrome is essential in gate designs, especially in the generalized control-X gates where only a single-photon loss on target qudit will induce a time-dependent phase shift on the control qudit. But the continuous syndrome monitoring is hard for multicomponent cat codes with stabilization. We will leave the discussion of qudit properties into further research.

Besides, instead of using continuous syndrome monitoring as we mentioned, we can also try to engineer jump operators $\hat a^\da \hat P_{\Delta+1}$ and $\hat b^\da \hat P_{\Delta-1}$ to achieve the autonomous error correction against single-photon loss~\cite{Albert2019}. It can give similar scaling results of the gate error probability while further reducing the overhead from feedback control. The details of this proposal are also worth to be worked out in further work.

In summary, we generalize the idea of construction of bias-preserving operations for cat code into pair-cat code to protect against a single-photon loss in either mode during gate operations. The continuous syndrome monitoring plays an essential role in the gate design to suppress errors. The generalization is quite straightforward due to the strong similarity between the two types of codes. Besides, the Hamiltonian protection of the pair-cat code is investigated and the large energy gap between code space and other states has also been found and numerically verified, which is another interesting feature in pair-cat code that is worth to explore in the future. 
% However, in this work we mainly focus on the regime that $|\gamma| \to +\infty$, the properties for pair-cat code working in finite $|\gamma|$ regime also need to be investigated. 

% However, in current setup photon gain error will still cause uncorrectable effects, which requires us to further develop bosonic codes that not only possess biased noise structure and a useful set of biased-preserving operations, but immune from both single-photon loss and gain error.
% Besides, the comparison of performance of pair-cat code with concatenated cat code, other stabilization and gate execution strategies and possible experimental realization methods of pair-cat code are worth to be explored in the future.

\begin{acknowledgements}

We acknowledge support from the ARO (W911NF-18-1-0020, W911NF-18-1-0212), ARO MURI (W911NF-16-1-0349, W911NF-21-1-0325), AFOSR MURI (FA9550-19-1-0399, FA9550-21-1-0209), AFRL (FA8649-21-P-0781), DoE Q-NEXT, NSF (OMA-1936118, EEC-1941583, OMA-2137642), NTT Research, and the Packard Foundation (2020-71479).

\end{acknowledgements}

\appendix

\section{Lower order Hamiltonian stabilization of pair-cat code}\label{AppLow}

In the main part, we have shown that the Hamiltonian $\hat H = -K(\hat a^{\da 2}\hat b^{\da 2} - \gamma^{*4})(\hat a^2 \hat b^2 - \gamma^4)$ can stabilize the pair-cat code space. Here we seek for Hamiltonian with lower maximum order $N$ that can also provide $\gamma$-dependent stabilization of pair-cat code space. Specifically, we hope to find the Hamiltonian
\begin{equation}
    \hat H = \sum_{m+n+p+q \leq N} f_{mn,pq} \hat a^{\da m}\hat a^n \hat b^{\da p}\hat b^q,
\end{equation}
such that both $|\gamma_{\Delta_0}\ra$ and $|(i\gamma)_{\Delta_0}\ra$ are eigenstates of $\hat H$ with the same eigen-energy. Without loss of generality we can specify the eigen-energy as $0$. So, we require
\begin{equation}\label{Stab}
    \hat H |\gamma_{\Delta_0}\ra = \hat H |(i\gamma)_{\Delta_0}\ra = 0.
\end{equation}

To make use of this requirement, we need to find a set of linearly independent states such that $\hat H |\gamma_{\Delta_0}\ra$ and $\hat H |(i\gamma)_{\Delta_0}\ra$ can be written as a linear superposition of them. We notice that $\{\hat P_\Delta \hat a^{\da n} \hat b^{\da m}|\gamma, \gamma\ra | n,m\in\mathbb{N}, \Delta \in \mathbb{Z}\}$ are linearly dependent due to the following identity:
\begin{equation}
\begin{split}
    &\hat P_\Delta \hat a^{\da n} \hat b^{\da m} |\gamma, \gamma\ra \\
    ={}& \hat P_\Delta \hat a^{\da (n+1)} \hat b^{\da (m-1)} |\gamma, \gamma\ra\\
    &+ \frac{\Delta + n - m + 1}{\gamma} \hat P_\Delta \hat a^{\da n} \hat b^{\da (m-1)} |\gamma, \gamma\ra.
\end{split}
\end{equation}
With this recursive formula, every $\hat P_\Delta \hat a^{\da n} \hat b^{\da m} |\gamma, \gamma\ra$ can be written as a linear superposition of $\{\hat P_\Delta \hat a^{\da k} |\gamma, \gamma\ra | k \leq n+m\}$. As a result, we can write $\hat H |\gamma_{\Delta_0}\ra$ as a linear superposition of $\hat P_\Delta \hat a^{\da k} |\gamma, \gamma\ra$ with different $k$ and $\Delta$, and Eq.~\eqref{Stab} requires all the coefficients are $0$. Similarly, we can write $\hat H |(i\gamma)_{\Delta_0}\ra$ under $\{\hat P_\Delta \hat a^{\da k} |i\gamma, i\gamma\ra | k \in \mathbb{N}; \Delta \in \mathbb{Z}\}$ to get another set of linear equations on $f_{mn,pq}$.

Besides, the Hermiticity of $\hat H$ requires that
\begin{equation}\label{Hermi}
    f_{mn, pq} = f^*_{nm, qp}.
\end{equation}

Eq.~\eqref{Stab} and \eqref{Hermi} form a set of linear equations of the real and imaginary part of $f_{mn, pq}$, and we hope to find all of its solutions with numerical help. In general, the solutions for $f_{mn, pq}$ should be a function of $\gamma$ and $\Delta_0$. However, even for one specific $\Delta_0$ we cannot find any $\gamma$-dependent solution up to $N = 6$. 

Here we call a set of solution $\gamma$-independent if (with proper overall factor since the set of linear equations is homogeneous) all of $f_{mn,pq}$ can be written as independent of $\gamma$. For example, when $\Delta_0 = 0$ we can easily find that $\hat \Delta = \hat b^\da \hat b - \hat a^\da \hat a$ is one solution since $\hat \Delta|\gamma_0\ra = \hat \Delta|(i\gamma)_0\ra = 0$, but all the non-zero coefficients of $\hat a^{\da m}\hat a^n \hat b^{\da p}\hat b^q$ in $\hat\Delta$ is either $-1$ or $1$, which are both independent of $\gamma$. So, we call this solution $\gamma$-independent. Other solutions that does not satisfy the $\gamma$-independent criteria are regarded as $\gamma$-dependent.

If we further restrict $\hat H$ to commute with all the $\hat \Delta$, which means the photon number difference of two modes is a conserved quantity, then all the non-zero $f_{mn, pq}$ should satisfy $n+p = m+q$, which means the maximum order in $\hat H$ should be an even number. So, our Hamiltonian $\hat H = -K(\hat a^{\da 2}\hat b^{\da 2} - \gamma^{*4})(\hat a^2 \hat b^2 - \gamma^4)$ with $N = 8$ is the one with lowest order that could provide both nontrivial $\gamma$-dependent protection and photon number difference conservation properties that we can find out.

\section{Structure of the stabilization Hamiltonian}\label{AppStru}

In this appendix, we briefly investigate the eigenstates and eigen-energies of the stabilization Hamiltonian $\hat H = -K(\hat a^{\da 2}\hat b^{\da 2} - \gamma^{*4})(\hat a^2 \hat b^2 - \gamma^4)$ defined in Eq.~\eqref{Hamil} in large $|\gamma|$ limit.

The first strategy is to perform displacement operation on the two modes. We denote $\hat D(\gamma, \gamma) = \hat D(\gamma) \otimes \hat D(\gamma)$ where $\hat D(\gamma)$ is the displacement operator. So, in the displaced frame of the two modes, the Hamiltonian is
\begin{equation}\label{Hdisp}
\begin{split}
    \hat H_{\disp, \gamma} &= \hat D(\gamma,\gamma)\hat H \hat D^\da(\gamma, \gamma)\\
    &= -8K|\gamma|^6\left(\frac{\hat a^\da + \hat b^\da}{\sqrt{2}}\right)\left(\frac{\hat a + \hat b}{\sqrt{2}}\right) + O(|\gamma|^5).
\end{split}
\end{equation}
% [LJ: In order for the expansion to converge, are we implicitly assuming that that $\hat{a},\hat{b} \ll \gamma$? If so, make a statement here.]
We can only keep the first term in $\hat H_{\disp, \gamma}$ if we just focus on the subspace where in the displaced frame the matrix elements of $\hat a$, $\hat b$ are far less than $\gamma$.

Besides, we can denote $\hat A = (\hat a + \hat b)/\sqrt{2}$ and $\hat B = (\hat a - \hat b)/\sqrt{2}$, which serve as two new independent modes. 

We define $|\psi\ra$ (with unit norm) as the ``asymptotic eigenstate" of an operator $\hat O(\gamma)$ if in the large $|\gamma|$ limit $|\psi\ra$ is parallel with $\hat O |\psi\ra$, or the norm of $\hat O |\psi\ra$ goes to zero in that limit. In this case, if we denote the states $|(n, m)\ra$ as
\begin{equation}
    |(n,m)\ra :=\frac{\hat A^{\da n} \hat B^{\da m}}{\sqrt{n!m!}}|0,0\ra,
\end{equation}
they will be the asymptotic eigenstates of $\hat H_{\disp, \gamma}$ in the large $|\gamma|$ limit as long as they satisfy either of the following two conditions. The one is $n \geq 1$ and $n, m \ll |\gamma|^2$, while the other one is $n = 0$ and $m = 0, 1$. So, in both of the two cases, $\hat D(\gamma, \gamma)|(n, m)\ra$ are the asymptotic eigenstates of $\hat H$. The derivation is also valid for states $\hat D(i\gamma, i\gamma) |(n,m)\ra$ which are also asymptotic eigenstates of $\hat H$ and long as $n, m$ satisfy the criteria we just mentioned.

It can also be shown that $|(0, m)\ra$ are not asymptotic eigenstates of $\hat H_{\disp, \gamma}$ when $m \geq 2$. We can calculate the angle between the two states $|(0,m)\ra$ and $\hat H_{\disp, \gamma}|(0,m)\ra$, and find out that they are not parallel but actually perpendicular to each other under large $|\gamma|$ limit when $m \geq 2$ because of the lower order corrections in Eq.~\eqref{Hdisp}.

In fact, this protocol can be generalized by using $\hat D(\gamma_1, \gamma_2)$ to find asymptotic eigenstates where $\gamma_1^2 \gamma_2^2 = \gamma^4$. The dominant part of the displaced Hamiltonian can be transformed into the form of a single oscillator via Gaussian operations. The energy spacing of the new mode is $4K|\gamma_1|^2|\gamma_2|^2 (|\gamma_1|^2 + |\gamma_2|^2)$, which is no less than $8K|\gamma|^6$ due to the constraint between $\gamma_1$ and $\gamma_2$.

\begin{figure}[t!]
\centering
\includegraphics[scale = 0.49]{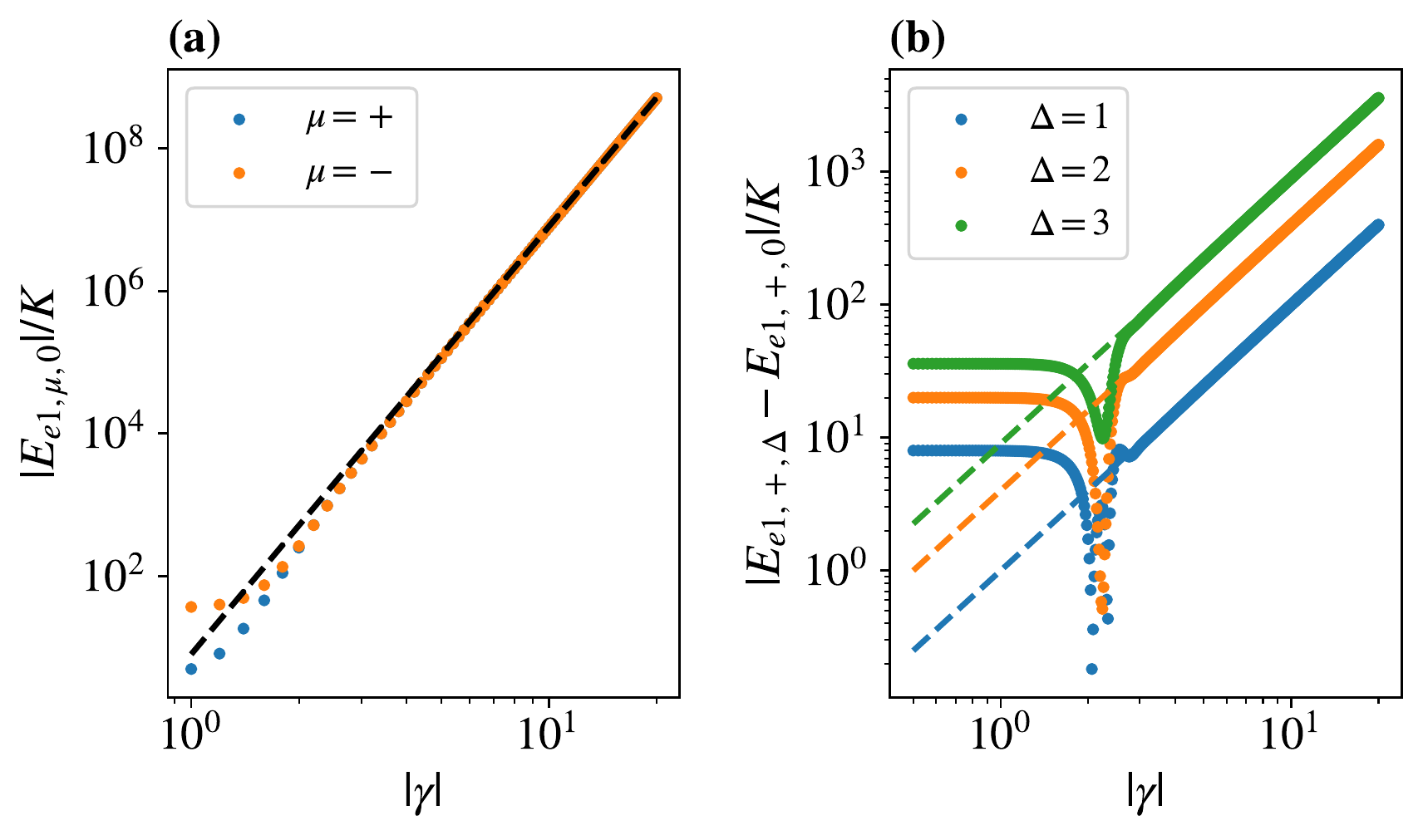}
\caption{(a) Energy gap $|E_{e1,\mu,0}|$ between code state $|\mu_{\gamma,\Delta}\ra$ and the ``first-less-excited" eigenstate in $\mu = +$ subspace and $\mu = -$ subspace. Dashed line corresponds to $|E_{e1,\mu,0}|/K = 8|\gamma|^6$. (b) Energy difference between $E_{e1,\mu,\Delta}$ and $E_{e1,\mu,0}$ with different $\Delta$. Dashed lines correspond to $|E_{e1,+,\Delta} - E_{e1,+,0}|/K = \Delta^2|\gamma|^2$, respectively.}
    \label{fig:EGap}
\end{figure}

To safely claim that the Hamiltonian in Eq.~\eqref{Hamil} can provide a protection of code space with the $8K|\gamma|^6$ energy gap, we also perform the exact diagonalization of the Hamiltonian with numerical help. We first separate $\hat H$ into different subspaces with fixed parity and photon number difference. Specifically, we focus on $\hat H_{\mu, \Delta}$ that mentioned in Eq.~\eqref{eq:Hsep}, and then numerically calculate the energy gap between $|\mu_{\gamma, \Delta}\ra$ state and the ``first-less-excited" eigenstate of $\hat H_{\mu, \Delta}$. In FIG.~\ref{fig:EGap}(a) we can see that in the $\Delta = 0$ case we do have $8K|\gamma|^6$ protection of the code space in large $|\gamma|$ limit.

In general, it is hard to write down the explicit form of all the asymptotic eigenstates of $\hat H_{\mu, \Delta}$, but we can see that the state $|\psi_{e1,\mu,\Delta}\ra$, which can be written as
\begin{equation}
    |\psi_{e1,\mu,\Delta}\ra \simeq \frac{\hat Q^{(\Delta)}_\mu (\hat a^\da + \hat b^\da - 2 \gamma^*) |\gamma,\gamma\ra}{\sqrt{2\N_{e1,\mu,\Delta}}},
\end{equation}
is the asymptotic eigenstate of $\hat H_{\mu, \Delta}$ with eigen-energy $E_{e1, \mu, \Delta} \simeq -8K|\gamma|^6$. Here $\N_{e1,\mu,\Delta}$ is a normalization factor. 

% With the explicit form of $|\psi_{e1,\mu,\Delta}\ra$, we can calculate the corresponding eigen-energy $E_{e1,\mu,\Delta} = \la\psi_{e1,\mu,\Delta}|\hat H|\psi_{e1,\mu,\Delta}\ra$ and the energy difference between different $\mu$ and $\Delta$. Numerically, we can see that all the $|\psi_{e1,\mu,\Delta}\ra$ states are approximately degenerate in large $|\gamma|$ limit for finite $\Delta$.

We can also numerically investigate the difference among $E_{e1, \mu, \Delta}$ for different $\mu$ and $\Delta$. Like the cat code case, $|E_{e1,+,\Delta} - E_{e1,-,\Delta}|$ is suppressed exponentially with $|\gamma|^2$ due to the exponentially suppressed overlap between any two states with large separation in the $\gamma$-plane, which means that the two ``first-less-excited" states with the same $\Delta$ and different parity are approximately degenerate in large $|\gamma|$ limit. Besides, $|E_{e1,\mu,\Delta} - E_{e1,\mu,0}| \sim O(\Delta^2 |\gamma|^2)$ in the large $|\gamma|$ limit with finite $\Delta$, which is a small correction compared with the $O(|\gamma|^6)$ gap. These facts together indicate that we do have the $O(|\gamma|^6)$ energy gap to protect the code space.

Those scaling results may change when considering another limit with finite $\gamma$ but focusing on the subspaces with $\Delta$ as large as possible. However, since typically we prefer to choose $\Delta = 0$ as the code space and the evolution is ideally $\Delta$-preserving, it is difficult for our states to go to the large $\Delta$ regime. So, we do not discuss this regime further but just point out this issue.

\section{Quantum error correction strategies of pair-cat code against photon loss}\label{AppQEC}

In this appendix, we will talk about the quantum error correction properties of the pair-cat code where noise only comes from photon loss. In the first part, we consider the system evolves under infinite strength of the dissipative stabilization while suffering from photon loss. We will talk about the recovery strategy and calculate the remaining error after the recovery process. In the second part, we consider the pair-cat code evolves with no stabilization but only photon loss, which means we can simulate the dynamics using a lossy bosonic channel (LBC). We will compare the results from the two cases. For simplicity, in this appendix we fix $\Delta = 0$ for our code space.

%In the first part, we do not apply any stabilization strategy but following the idea in Ref.~\cite{Li2017} to focus on the outcomes of the pair-cat code passing through a lossy bosonic channel (LBC) for each mode and the corresponding recovery process. In the second part, we will consider another case where infinite strength of the dissipative stabilization has been turned on during the lossy process. 

\subsection{Lossy process with dissipative stabilization}

In this part, we consider the situation where the dissipative stabilization has been turned on during the lossy process. Specifically, the evolution channel can be written as
\begin{equation}
    \mathcal{E}^D_\evo \rho = e^{\mathcal{L}t} \rho = e^{\kappa t \mathcal{L}_D + \kappa_1 t \mathcal{L}_E} \rho,
\end{equation}
where $\mathcal{L}_D = \mathcal{D}[\hat a^2 \hat b^2 - \gamma^4]$ and $\mathcal{L}_E = \mathcal{D}[\hat a] + \mathcal{D}[\hat b]$. To simplify the derivation, we will consider the extreme situation where $\kappa \to +\infty$, such that for any $t>0$ we have $e^{\kappa t \mathcal{L}_D} \rho = \mathcal{P}\rho = \hat P_D \rho \hat P_D$, where $\hat P_D = \sum_{\mu, \Delta} |\mu_{\gamma, \Delta}\ra\la \mu_{\gamma, \Delta}|$ is the projection operator for the subspace stabilized by the dissipator $\mathcal{D}[\hat a^2 \hat b^2 - \gamma^4]$. Then, because of the following identity~\cite{Lebreuilly2021}:
\begin{equation}
    \begin{split}
        e^{(A+B)t} ={}& e^{At} + \int_0^t \dif s \ e^{(A+B)(t-s)} B e^{As}\\
        ={}& e^{At} + \sum_{k=1}^N \ \iint\limits_{\sum_i \tau_i \leq t}(\prod_{i=1}^k \dif \tau_i) e^{A(t - \sum_i \tau_i)} \prod_{i=1}^k (Be^{A\tau_i})\\
        &+\iint\limits_{\sum_i \tau_i \leq t}(\prod_{i=1}^{N+1} \dif \tau_i) e^{(A+B)(t - \sum_i \tau_i)} \prod_{i=1}^{N+1} (Be^{A\tau_i}),
    \end{split}
\end{equation}
we have
\begin{equation}
    \mathcal{E}^D_\evo = \mathcal{P} + \sum_{k=1}^N \frac{(\kappa_1 t)^k}{k!} (\mathcal{P} \mathcal{L}_E \mathcal{P})^k + O[(\kappa_1 t)^{N+1}].
\end{equation}

% \ming{[This part has been re-written. The recovery process might not be optimal but intuitively correct. I believe that in the case where $\epsilon$ or $\gamma$ is large so that $\hat a\hat b$ is more likely than nothing happens, or $\epsilon_a \neq \epsilon_b$, there will be better recovery strategies.]} 
Now let us discuss the recovery process. As mentioned in the main text, we should first measure $\hat \Delta$ of the final states to extract the syndrome and then decide which operation we should apply. One intuitive way is to assume all the loss errors happen only in one mode, since other loss errors that lead to the same $\Delta$ correspond to higher order of $\kappa_1$, which are less likely to happen in the case that $\kappa_1 \gamma^2 t \ll 1$. So, if the final $\Delta > 0$ we will assume that loss only happens in mode $\hat a$, and if $\Delta < 0$ we assume that loss only happens in mode $\hat b$.

We notice that the code states $|\mu_{\gamma, \Delta}\ra$ defined in Eq.~\eqref{codestates} satisfy
\begin{equation}
\hat a^k \hat b^l |\mu_{\gamma, 0}\ra = \gamma^{k+l} \sqrt{\frac{\N_{\mu', k-l}}{\N_{\mu, 0}}}|\mu'_{\gamma, k-l}\ra,
\end{equation}
where
\begin{equation}\label{mu_pr}
\mu' = \mu \cdot (-1)^{\max(k,l)}.
\end{equation}
Here we use $\mu=+1$ to indicate even ($+$) parity and $\mu=-1$ for odd ($-$) parity. It is also worth to point out here that for $\Delta < 0$ case $\N_{\mu, \Delta} := \N_{\mu, |\Delta|}$.

As a result, if the final $\Delta = k - l$ is odd, we will assume either $k$ or $l$ is zero and another is odd, so according to Eq.~\eqref{mu_pr} $\mu'$ is different from $\mu$ and recover operation should be able to restore the parity of the states; and if $\Delta$ is even, we assume $\mu'$ does not change from $\mu$. It is easy to see that $\hat a\hat b$ will result in an uncorrectable error under this strategy, because in this case the final $\Delta = 0$ is even and we will assume $\mu' = \mu$. However, according to Eq.~\eqref{mu_pr} $\mu'$ is different from $\mu$ since $k = l = 1$, which means our assumption is wrong and it will cause an error even after the recovery process. Besides, the amplitude $\gamma$ does not change after evolution due to the strong stabilization we use.
  
In summary, the recovery channel $\mathcal{R}^D$ can be chosen as a set of $\Delta$-dependent unitary operations $\hat R_\Delta = \sum_{\mu'=\pm} |\mu''_{\gamma, 0}\ra \la\mu'_{\gamma, \Delta}|$ that map $|\mu'_{\gamma, \Delta}\ra$ to $|\mu''_{\gamma, 0}\ra$, where $\mu'' =\mu' \cdot (-1)^{|\Delta|}$. So, we can write
\begin{equation}
    \mathcal{R}^D\rho = \sum_{\Delta} \hat R_\Delta\rho\hat R^\da_\Delta.
\end{equation}

\begin{figure}[t!]
\centering
\includegraphics[scale = 0.49]{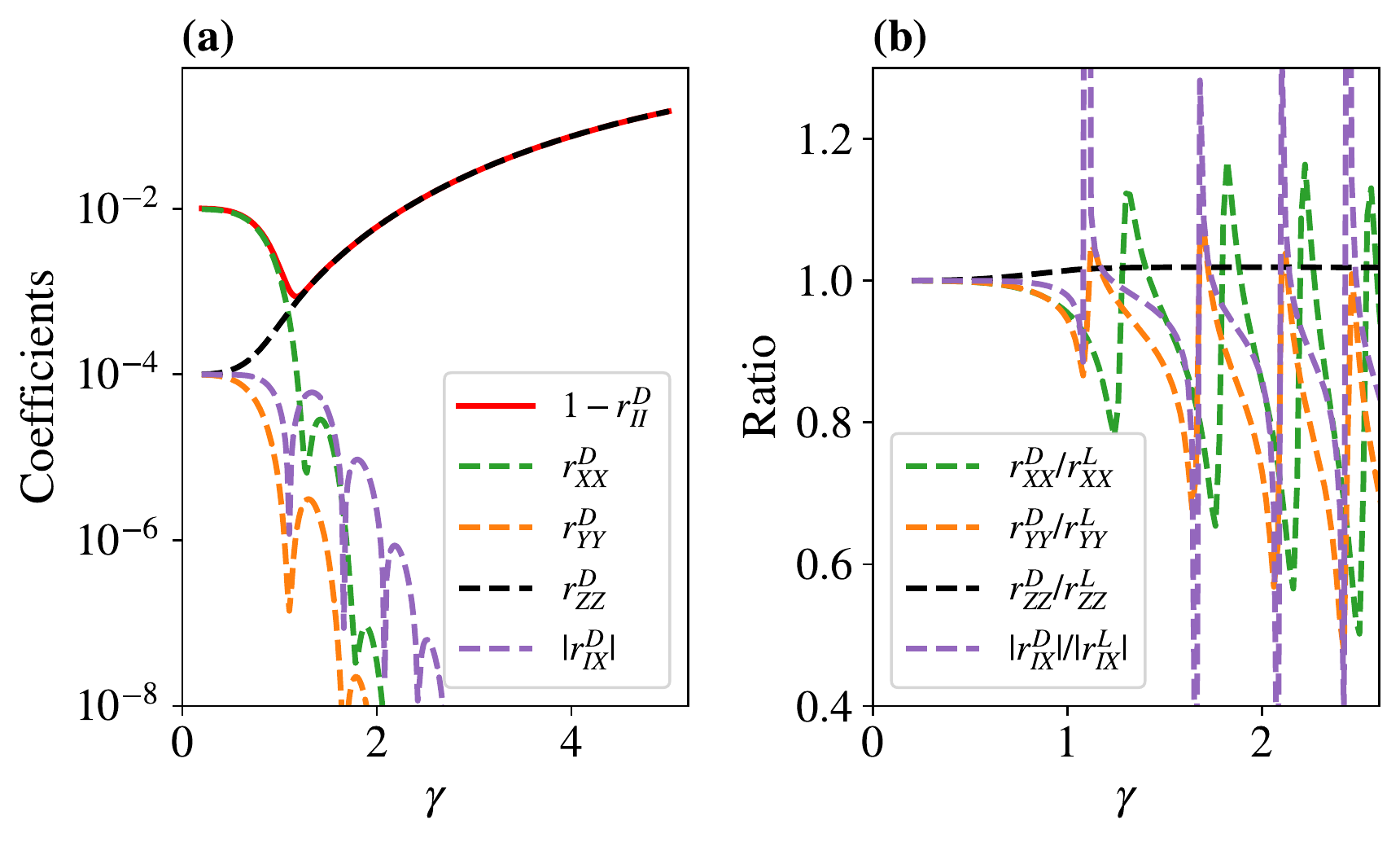}
\caption{(a) The coefficients of the process tomography of $\mathcal{E}^D_\tot$ with $1 - e^{-\kappa_1 t} = 2\%$. (b) Comparison between $|r^{D}_{jk}|$ and $|r^L_{jk}|$ with $\epsilon = 1 - e^{-\kappa_1 t} = 2\%$.}
    \label{fig:lossy_channel}
\end{figure}

Finally we can investigate the effect of whole process $\mathcal{E}^D_\tot = \mathcal{R}^D \circ \mathcal{E}^D_\evo$ acting on $\rho$ that lies in the code space spanned by $|\pm\ra_c = |\pm_{\gamma, 0}\ra$. With numerical help, we can calculate the coefficients of the process tomography under Pauli basis to indicate the error probability after the recovery channel. Specifically, we have
\begin{equation}\label{eq:pro_tomo}
    \mathcal{E}^D_\tot\rho = \sum_{j, k \in \{I,X,Y,Z\}} r^D_{jk} \hat W_j \rho \hat W^\da_k,
\end{equation}
\color{red}
%[LJ: Could you check numerically if we can approximate this expression using only diagonal terms? That is, $\mathcal{E}^D_\tot\rho \approx \sum_{k,k \in \{I,X,Y,Z\}} r^D_{kk} \hat W_k \rho \hat W^\da_k$. If it is true, then we only need to show the coefficients $r^D_{kk}$.]
\color{black}
where $\hat W_j \in \{\hat I_c, \hat X_c, \hat Y_c, \hat Z_c\}$. In FIG.~\ref{fig:lossy_channel}(a), from the diagonal term $r^D_{jj}$ we can clearly see the bias structure of the noise and find the local optimal $\gamma$ value to suppress the bit-flip error, or the total error itself. For the off-diagonal term of $r^D_{jk}$, numerically we can see that only $r^D_{IX(XI)}$ and $r^D_{YZ(ZY)}$ are non-zero, and we can also find $|r^D_{IX}| = |r^D_{YZ}|$. In the small $\gamma$ regime, we have $|r^D_{IX}| \ll r^D_{XX}$, while when $\gamma$ is large we have $|r^D_{IX}| \ll r^D_{ZZ}$.

Here we provide a simple explanation for the origin of those error terms. $\hat X_c$ term comes from the slight difference of the photon loss error probability between $|+\ra_c$ and $|-\ra_c$ as shown in Eq.~\eqref{eq:dephasing}, which decreases exponentially as the increase of $|\gamma|^2$. $\hat Z_c$ and $\hat Y_c$ terms come from the uncorrectable photon loss $\hat a\hat b$, and according to Eq.~\eqref{eq:abloss} $Z$ error will increase as $|\gamma|$ becomes larger, while the $\hat Y_c$ term will be suppressed exponentially.

In the $\gamma \to 0$ limit, we have $|+\ra_c = |0,0\ra$ and $|-\ra_c = |1,1\ra$ under Fock basis. $\hat P_D$ projects into the subspace spanned by those Fock states where either $\hat a$ or $\hat b$ mode is populated with at most one photon. We can keep the terms up to $O[(\kappa_1 t)^2]$ in $\mathcal{E}^D_\evo$ and then derive $\mathcal{E}^D_\tot$. It is easy to see that $r^D_{XX} \sim O(\kappa_1 t)$ while $r^D_{ZZ}$, $r^D_{YY}$ and $|r^D_{IX(YZ)}|$ scales as $O[(\kappa_1 t)^2]$, which indicates that $X$ error is the dominant one in the $\gamma \to 0$ limit.

%For the recovery process after evolution, we still first measure $\hat \Delta$ and then apply the $\Delta$-dependent unitary operation $\hat R_\Delta = \sum_{\mu'=\pm} |\mu''_{\gamma, 0}\ra \la\mu'_{\gamma, \Delta}|$ where $\mu'' =\mu' \cdot (-1)^{|\Delta|}$, which is similar to the strategy we mentioned in the no stabilization case, except for the fact that due to the stabilization the amplitude $\gamma$ does not change during evolution. We can also calculate the process tomography of $\mathcal{E}_\tot = \mathcal{R} \circ \mathcal{E}_\evo$ as we did in Eq.~\eqref{eq:pro_tomo} and plot the $\gamma$-dependence of those coefficients in FIG.~\ref{fig:lossy_channel}(b). We can find similar behavior with the no stabilization case when $\gamma$ is relatively large.

% We can see that, similar to the cat code results shown in Ref.~\cite{Li2017}, in the small $\epsilon$ and large $|\gamma|$ regime, the full channel $\mathcal{E}$ can be approximated as the following form of a Pauli channel:
%\begin{equation}
%    \mathcal{E}_\tot\rho \approx (1 - r_{XX} - r_{ZZ}) \hat I_c\rho \hat I_c + r_{XX} \hat X_c \rho \hat X_c + r_{ZZ} \hat Z_c \rho \hat Z_c.
%\end{equation}

\subsection{Lossy bosonic channels}

In this part, we consider another situation that there is no stabilization but only photon loss during evolution. Therefore, we can treat the evolution process as an LBC, whose effects on cat code have been discussed in Ref.~\cite{Li2017}. We will first introduce what the LBC is, and then develop the recovery protocol and finally estimate the remaining errors. We consider that both $\hat a$ mode and $\hat b$ mode suffer independently from an LBC which can be written in the following form~\cite{Chuang1997}:
\begin{equation}
    \mathcal{E}^L_\evo\rho = \sum_{k, l = 0}^{+\infty} \hat E^a_k \hat E^b_l \rho \hat E^{a \da}_k \hat E^{b \da}_l,
\end{equation}
where $\hat E^a_k := (k!)^{-1/2} \epsilon_a^{k/2} (1-\epsilon_a)^{\hat n_a / 2}\hat a^k$ and $\hat E^b_l$ can be written in the same way for the $\hat b$ mode. For simplicity we assume $\epsilon_a = \epsilon_b = \epsilon$. In cavities with photon loss rate $\kappa_1$ and evolution time $t$, we have $\epsilon = 1 - e^{-\kappa_1 t}$. Then we will fix $\Delta = 0$ as our code space and the recover strategy is based on the final $\Delta$ we get after the lossy channel. After being applied by the Kraus operators of LBC, the code state $|\mu_{\gamma, \Delta}\ra$ will become:
\begin{equation}
    \hat E^a_k \hat E^b_l |\mu_{\gamma, 0}\ra = \frac{(\gamma\sqrt{\epsilon} )^{k+l}}{\sqrt{k!l!}}e^{-\epsilon |\gamma|^2} \sqrt{\frac{\N_{\mu', k-l}^{\gamma'}}{\N_{\mu, 0}^\gamma}} |\mu'_{\gamma', k-l}\ra,
\end{equation}
where $\N_{\mu, \Delta}^\gamma$ is value of the normalization function $\N_{\mu, \Delta}$ at $\gamma$. Besides, we have $\gamma' = \gamma\sqrt{1-\epsilon}$ and $\mu'$ is the same as that in Eq.~\eqref{mu_pr}.

For the recovery process after evolution, we still first measure $\hat \Delta$ and then apply the $\Delta$-dependent unitary operation $\hat R_\Delta = \sum_{\mu'=\pm} |\mu''_{\gamma, 0}\ra \la\mu'_{\gamma', \Delta}|$ where $\mu'' =\mu' \cdot (-1)^{|\Delta|}$, which is similar to the strategy we mentioned in the strong dissipative stabilization case except for the fact that without stabilization we have $\gamma' \neq \gamma$. Here we can explicitly write down the matrix representation of $\mathcal{E}^L_\tot = \mathcal{R}^L \circ \mathcal{E}^L_\evo$ under $\rho = \begin{pmatrix} \rho_{++}& \rho_{+-} & \rho_{-+} & \rho_{--}\end{pmatrix}^\text{T}$ where $\rho$ is written under the $|\pm\ra_c = |\pm_{\gamma, 0}\ra$ basis. Denote $\mathcal{M}_{k,l} = (\epsilon|\gamma|^2)^{k+l}e^{-2\epsilon |\gamma|^2}/(k!l!)$, we have
\begin{widetext}
\begin{equation}\label{eq:EpsiL}
\mathcal{E}^L_\tot =
\sum_{k, l = 0}^{+\infty} \mathcal{M}_{k,l} \begin{pmatrix} \delta_{\min(k,l),0}^{\text{mod } 2}\frac{\N_{\mu_+, k-l}^{\gamma'}}{\N_{+, 0}^\gamma} & 0 & 0 & \delta_{\min(k,l),1}^{\text{mod } 2}\frac{\N_{\mu_-, k-l}^{\gamma'}}{\N_{-, 0}^\gamma} \\
0 & \delta_{\min(k,l),0}^{\text{mod } 2}\sqrt{\frac{\N_{+, k-l}^{\gamma'}\N_{-, k-l}^{\gamma'}}{\N_{+, 0}^\gamma\N_{-, 0}^\gamma}} & \delta_{\min(k,l),1}^{\text{mod } 2}\sqrt{\frac{\N_{+, k-l}^{\gamma'}\N_{-, k-l}^{\gamma'}}{\N_{+, 0}^\gamma\N_{-, 0}^\gamma}} & 0\\
0 & \delta_{\min(k,l),1}^{\text{mod } 2}\sqrt{\frac{\N_{+, k-l}^{\gamma'}\N_{-, k-l}^{\gamma'}}{\N_{+, 0}^\gamma\N_{-, 0}^\gamma}} &\delta_{\min(k,l),0}^{\text{mod } 2}\sqrt{\frac{\N_{+, k-l}^{\gamma'}\N_{-, k-l}^{\gamma'}}{\N_{+, 0}^\gamma\N_{-, 0}^\gamma}} & 0\\
\delta_{\min(k,l),1}^{\text{mod } 2}\frac{\N_{\mu_+, k-l}^{\gamma'}}{\N_{+, 0}^\gamma} & 0 & 0 & \delta_{\min(k,l),0}^{\text{mod } 2}\frac{\N_{\mu_-, k-l}^{\gamma'}}{\N_{-, 0}^\gamma} \\
\end{pmatrix},
\end{equation}
\end{widetext}
where $\delta_{j, k}^{\text{mod } 2} = 1$ if $(j-k)$ is even, and $\delta_{j, k}^{\text{mod } 2} = 0$ otherwise. Besides, $\mu_\pm := (\pm 1) \cdot (-1)^{\max(k,l)}$. We can also find the coefficients of process tomography and denote them as $r^L_{jk}$. With the explicit form of $\mathcal{E}_\tot^L$ in Eq.~\eqref{eq:EpsiL}, it is easy to check that $r^L_{IY} = r^L_{IZ} =r^L_{XY} =r^L_{XZ} = 0$ and $|r^L_{IX}| = |r^L_{YZ}|$. In FIG.~\ref{fig:lossy_channel}(b), we compare the results between $r^D_{jk}$ and $r^L_{jk}$ under $\epsilon = 1 - e^{-\kappa_1 t} = 2\%$. We can see that $r_{ZZ}$ is approximately the same between the two situations. However, since the local minima in $r^D_{XX}(\gamma)$ and $r^L_{XX}(\gamma)$ correspond to slightly different $\gamma$, the $r^D_{XX}/ r^L_{XX}$ curves fluctuate near $1$. This argument also works for the $r_{YY}$ and $r_{IX}$ cases.

\section{Perturbative analysis of gate errors}\label{AppError}

In this appendix, we will discuss about the scaling of the Z error probability of the pair-cat code during gate operations in the dissipative stabilization scheme. This type of error can be induced by both photon loss and the leakage out of the protected code space in the middle of the gate execution. Since photon loss errors will not cause leakage out of the stabilized subspace, we will treat the two effects separately. Our analysis is based on the methods introduced in~\cite{Chamberland2022} where gate errors are investigated for cat code. There adiabatic elimination method~\cite{Reiter2012} has been used in order to achieve effective dynamical equations in the stabilized subspace. Due to the similarities between cat code and pair-cat code, we can also use similar strategies to derive error probability of the gates we construct for the pair-cat code. Therefore, in the following discussion we will only mention those key ingredients in the derivation to achieve $Z$ error probability and properties specialized for pair-cat code, while detailed reasoning for each step of the derivation can be found in~\cite{Chamberland2022} that focuses on the cat code counterpart.
We highlight that for the pair-cat code all the gate error probability can be achieved to scale at least linearly in single-photon loss rate $\kappa_1$, which works better compared with cat code where error probability mainly scales as $\sqrt{\kappa_1}$~\cite{Guillaud2019,Chamberland2022}. To simplify the notation, we will assume $\gamma$ to be a positive real number.

\subsection{$Z(\theta)$ gate}

The $Z(\theta)$ gate can be implemented via the 2-mode squeezing Hamiltonian $\hat H_Z$ in Eq.~\eqref{HZHZZ}. The $\hat a^\da \hat b^\da$ term will cause leakage out of the code space via
\begin{equation}\label{adbd}
    \hat a^\da \hat b^\da |\mu_{\gamma, \Delta}\ra \simeq  \gamma^2 |\mu'_{\gamma, \Delta}\ra + \sqrt{2}\gamma |\psi_{e1,\mu',\Delta}\ra + O(1),
\end{equation}
where $\mu' \neq \mu$. It means the parity of the two modes are changed. Further, this leakage can be recovered back to the code space via the stabilization dissipator $\kappa \mathcal{D}[\hat a^2\hat b^2 - \gamma^4]$, since
\begin{equation}
    (\hat a^2\hat b^2 - \gamma^4)|\psi_{e1,\mu',\Delta}\ra \simeq 2\sqrt{2}\gamma^3 |\mu'_{\gamma, \Delta}\ra.
\end{equation}
This whole process together will give us a $Z$ error on the pair-cat code. The effective error rate can be estimated via adiabatic elimination method, which will give us $4(\epsilon_Z \sqrt{2} \gamma)^2/(8\kappa\gamma^6) = \epsilon_Z^2/(\kappa \gamma^4)$.

Z error probability induced by photon loss is simple to be analyzed. Only a single-photon loss in one mode will not cause errors in the pair-cat code since it can be detected by $\hat \Delta$ measurement at the end of the gate operation and correct it via a recovery channel. $Z$ error will be induced if both $\hat a$ and $\hat b$ errors happens. Therefore, the combined error probability is $(\kappa_1 \gamma^2 T)^2$, and the total dephasing error probability during the $Z(\theta)$ gate is
\begin{equation}
    p_Z = \frac{\epsilon_Z^2}{\kappa \gamma^4}T + (\kappa_1 \gamma^2 T)^2
\end{equation}
Recall that $\epsilon_Z = -\theta/(4\gamma^2 T)$, we can find the optimal time $T$ to minimize $p_Z$ scales as $T^\opt = (\theta^2/32\kappa\kappa_1^2\gamma^{12})^{1/3}$, and the corresponding $p^\opt_Z \sim O((\kappa_1/\kappa)^{2/3}/\gamma^4)$.

On the other hand, this scaling can be changed by using real-time monitoring of $\hat \Delta$ as we introduced in the design of CNOT gate. Suppose we keep measuring $\hat \Delta$ at a time interval of $\delta \tau$, then only the case that both $\hat a$ and $\hat b$ errors happen within the $\delta \tau$ interval will cause Z error, otherwise we will know exactly that the code does suffer from $\hat a\hat b$ loss instead of nothing happens. In this case, the $Z$ error probability induced by loss is $(\kappa_1 \gamma^2 \delta\tau)^2(T/\delta \tau) = \kappa_1^2 \gamma^4 \delta \tau T$, and the optimal $p^\opt_Z \sim O(\kappa_1 \sqrt{\delta \tau / \kappa}/\gamma^2)$, which is linear in $\kappa_1$.

\subsection{$ZZ(\theta)$ gate}

Like the $Z(\theta)$ gate, the Hamiltonian $\hat H_{ZZ}$ in Eq.~\eqref{HZHZZ} that we use to construct $ZZ(\theta)$ gate will also cause leakage out of the code space, which will induce $Z$ type of errors. For example, the $\hat a_1\hat b_1 \hat a^\da_2 \hat b^\da_2$ term couples the code state of the second qubit with states out of the code space while changing the parities of both pair-cat qubits at a rate of $\sqrt{2}\epsilon_{ZZ} \gamma^3$, which is the dominant leakage rate from this term. Besides, $\hat a^\da_1\hat b^\da_1 \hat a_2 \hat b_2$ term has similar effect but causes leakage in the first qubit. These two channels, after going back to the code space due to the dissipative stabilization, will cause a $Z_1Z_2$ error on the pair-cat qubits at a rate of $2\times 4(\sqrt{2}\epsilon_{ZZ}\gamma^3)^2/(8\kappa\gamma^6) = 2\epsilon_{ZZ}^2/\kappa$. So,
\begin{equation}
    p_{Z_1Z_2} = 2\epsilon_{ZZ}^2 T/\kappa = \frac{\theta^2}{8\kappa \gamma^8 T}
\end{equation}

The photon loss induced dephasing error can be analyzed in the same way as the former case in $Z(\theta)$ case. If there is no real-time $\hat \Delta$ monitoring, then $p_{Z_1} = p_{Z_2} = (\kappa_1 \gamma^2 T)^2$, and the optimal total error probability $p = p_{Z_1} + p_{Z_2} + p_{Z_1Z_2}$ scales as $O((\kappa_1/\kappa)^{2/3}/\gamma^4)$. If we have this real-time $\hat \Delta$ monitoring with time interval $\delta \tau$, then $p_{Z_1} = p_{Z_2} = \kappa_1^2 \gamma^4 \delta \tau T$ and optimal $p$ scales as $O(\kappa_1 \sqrt{\delta \tau / \kappa}/\gamma^2)$.

\subsection{$X$ gate}

The $X$ gate is implemented by changing stabilization parameter $\gamma$ with respect to time that $\gamma(t) = \gamma e^{i\frac{\pi}{2}\frac{t}{T}}$ and use an extra Hamiltonian $\hat H_{X, \text{rot}} = -\frac{\pi}{2T}(\hat n_a + \hat n_b)$ to compensate the error induced by non-adiabaticity. If there is no loss happening during this gate execution, the $X$ gate can be implemented perfectly, and there is no term to cause leakage out of the code space. So, the only source that can induce dephasing error is from photon loss.

Compared with the idling case where $\gamma$ stays as a constant, a single-photon loss at time $t_0$ in either mode will induce an extra global phase $e^{i\frac{\pi}{2}\frac{t_0}{T}}$. Unlike in the CNOT gate case we mentioned in the main text where this induced phase on the target qubit does cause a $Z$ type rotation of the control qubit, here this is just a global phase on a single qubit that does not matter at all. As a result, Z error probability induced by photon loss during $X$ gate has no difference compared with other gates: if there is no real-time $\hat \Delta$ monitoring, $p_Z = (\kappa_1 \gamma^2 T)^2$; if we have such monitoring, $p_Z = \kappa_1^2 \gamma^4 \delta \tau T$. So, to reduce $p_Z$, we should choose $T$ to be as small as possible.

\subsection{CNOT gate}
The CNOT gate can be implemented by changing stabilization parameter $\gamma(t)$ of the target qubit conditioned on the states of the control qubit via jump operators $\hat F_{1,2}$ defined in Eq.~\eqref{FCNOTpc}. We also apply another Hamiltonian $\hat H_{\CNOT, \text{rot}}$ in Eq.~\eqref{HCNOTpcrot} to reduce the error induced by non-adiabaticity. However, this extra Hamiltonian cannot fully compensate it, because other than the desired conditional rotation $\hat H' = -\frac{\pi}{2T} |1\ra_{1c}\la 1|\otimes (\hat a_2^\da \hat a_2 + \hat b_2^\da \hat b_2 - 2|\gamma|^2)$, the $\hat a_1^\da \hat b_1^\da (\hat a_2^\da \hat a_2 + \hat b_2^\da \hat b_2 - 2|\gamma|^2)$ term can also cause excitation of both control and target qubits. Other than Eq. ~\eqref{adbd}, we also have
\begin{equation}
    (\hat a^\da \hat a + \hat b^\da \hat b - 2|\gamma|^2)|\mu_{\gamma,\Delta}\ra \simeq \sqrt{2}\gamma |\psi_{e1, \mu, \Delta}\ra,
\end{equation}
where the parity of the target qubit does not change in this process. But, as discussed before, the parity of control qubit will change, which effectively cause a $Z$ operation on it.

We follow the method in Ref.~\cite{Chamberland2022} by going to the rotating frame according to $\hat H'$. In this frame, $\hat a_2 \hat b_2$ will be transformed to
\begin{equation}\label{rot_decay}
    \hat a_2 \hat b_2 \to |0\ra_{1c}\la 0| \otimes \hat a_2 \hat b_2 + |1\ra_{1c}\la 1| \otimes \hat a_2 \hat b_2 e^{i\pi t/T}
\end{equation}
This will result that, if the both control and target qubit state get excited together as we mentioned and then decay back to the code space via $\hat F_1$ and $\hat F_2$, effectively there will be a $Z_1Z_1(-\frac{2\pi t}{T}) = Z_1(\pi\frac{T-2t}{T})$ error acting on the control qubit. It is because that in the rotating frame going back to the code space of target qubit via $\hat F_2$ will cause a $Z_1(-\frac{2\pi t}{T})$ operation on the control qubit. The corresponding effective error rate is $\frac{\pi^2}{64\kappa \gamma^6 T^2}$. This $Z_1(\theta)$ error does not change when going back to the original frame, and after averaging with total time the $Z_1$ error probability induced from the non-adiabaticity is $\frac{\pi^2}{128\kappa \gamma^6 T}$. It is also worth to mention that in the derivation of the error probability we have ignored the situation that the $\hat a_1 \hat b_1$ term in $\hat F_2$ can also help the control qubit to decay back to the code space, but this term will not cause the change of the scaling property of the result we achieved, as discussed in~\cite{Chamberland2022}.

Then let us discuss the error induced by photon loss. We assume to have real-time $\hat \Delta$ monitoring on both control and target qubits. The photon loss on control qubit does not affect the phase of the target qubit state, so it will just give a $Z_1$ error with probability $\kappa_1^2 \gamma^4 \delta \tau T$. Things will become different when loss errors happen on target qubit. As discussed in the main text, the photon loss on target qubit will induce a time-dependent phase shift on control qubit, therefore we want to use real-time $\hat \Delta$ measurement to monitor when the loss happens and correct it with an extra $Z_1(\theta)$ gate. 

In practice, however, there are two relevant processes to induce $Z$ type of errors because of photon loss on target qubit. One comes from that, even only a single-photon loss happens, due to the finite $\delta \tau$ time of two $\Delta$ measurements, we can only correct the extra phase for $|1\ra_{1c}$ state of the control qubit up to small deviation ranging within $[-\frac{\delta \theta}{2}, \frac{\delta \theta}{2}]$, where $\delta \theta = \frac{\pi}{2} \frac{\delta \tau}{T}$. This effect on average will give a $Z_1$ error probability as $\frac{\kappa_1\gamma^2 (\delta \tau)^2 \pi^2}{96T}$. The other comes from that both $\hat a_2$ and $\hat b_2$ happen to the target qubit within $\delta \tau$ time, which not only causes a $Z_2$ error on target qubit but also induces an extra $e^{i\pi t/T}$ phase on $|1\ra_{1c}$ state of the control qubit. Using the same method we did in Eq.~\eqref{qjump}, we found that this error has the form of $Z_1Z_1(-\frac{\pi t}{T}) Z_2 = Z_1(\pi\frac{T-t}{T}) Z_2$ 
% \color{red}
% [LJ: Shall we write $Z_1(\pi\frac{T-t}{T}) Z_2$?]
% \color{black}
% \ming{[MY: I prefer to keep the similar notation as \cite{Chamberland2022}]}
with error rate $\kappa_1^2 \gamma^4 \delta \tau$, which on average gives both $Z_2$ error and $Z_1Z_2$ error a probability of $\kappa_1^2 \gamma^4 \delta \tau T / 2$.

In summary, we have 
\begin{equation}
    \begin{split}
        & p_{Z_1} = \kappa_1^2 \gamma^4 \delta \tau T + \frac{\pi^2}{128\kappa \gamma^6 T} + \frac{\kappa_1 \gamma^2 (\delta \tau)^2 \pi^2}{96T},\\
        & p_{Z_2} = p_{Z_1Z_2} = \frac{\kappa_1^2 \gamma^4 \delta \tau T}{2}.
    \end{split}
\end{equation}
By choosing the optimal time $T$ to minimize the total error probability $p = p_{Z_1} + p_{Z_2} + p_{Z_1Z_2}$, we find that $p \sim O(\sqrt{\frac{\kappa_1}{\kappa}} \sqrt{\frac{\kappa_1 \delta\tau}{\gamma^2}}\sqrt{ 1 + C \kappa\kappa_1\gamma^8 (\delta\tau)^2})$, where $C$ is a constant. Therefore, the total error probability scales between $O(\kappa_1)$ and $O(\kappa_1^{3/2})$.

\subsection{Toffoli gate}

The error properties in Toffoli gate are similar with those in CNOT gate and we can use the same method to analyze them. For convenience we denote $CZ(\theta) := \exp(-i\theta |11\ra\la 11|)$. Recall the Hamiltonian defined in Eq.~\eqref{HTofpcrot} which is used to compensate the non-adiabatic error, it could cause joint excitation of 1, 3 states or 2, 3 states out of their code spaces, together with a parity change in either qubit 1 or qubit 2 that gets excited. Again, by going to the rotating frame according to $\hat H' = -\frac{\pi}{2T} |1\ra_{1c}\la 1|\otimes |1\ra_{2c}\la 1|\otimes (\hat a_3^\da \hat a_3 + \hat b_3^\da \hat b_3 - 2|\gamma|^2)$, we realize that the two control qubit states will gain a phase if they are in $|1\ra_{1c}|1\ra_{2c}$ when the target qubit state decays back to the code space due to the dissipator $\hat F_3$ defined in Eq.~\eqref{FTofpc3}. This is similar to the effect caused by Eq.~\eqref{rot_decay} in the CNOT gate, but here we need to focus on the transformation of $\hat a_3 \hat b_3$ into the rotating frame instead. By using adiabatic elimination method, the effective error will be either $\frac{I-Z_1}{2}Z_2 CZ_{1,2}(-\frac{2\pi t}{T})$ or $\frac{I-Z_2}{2}Z_1 CZ_{1,2}(-\frac{2\pi t}{T})$ with the same error rate $\pi^2/(64\kappa\gamma^6 T^2)$. Going back to the original frame will not cause the change of the error forms. After averaging over time, we can see the non-adiabaticity will give an error probability of $\frac{\pi^2}{256\kappa\gamma^6 T}$ for all $Z_1$, $Z_2$, and $Z_1Z_2$ type of errors. Similar to the derivation in the CNOT case, we also ignore the contribution that the excitations of two control qubits can decay back via $\hat F_3$ since this effect does not change the scaling of the error probability we derived.

Then we discuss about $Z$ type of error induced by photon loss. We again assume to have real-time $\hat \Delta$ monitoring of all three pair-cat qubits. The loss happens in either of the two control qubits will not cause errors in qubits that do not suffer from photon loss, so only both $\hat a_i$ and $\hat b_i$ ($i = 1, 2$) happen in the same time interval of $\hat \Delta$ measurement will cause a $Z_i$ error. The corresponding error probability is again $\kappa_1^2 \gamma^4 \delta \tau T$.

Similar to the CNOT gate, the loss errors happen in the target qubit will also cause two effects. First, a single-photon loss of either $\hat a_3$ or $\hat b_3$ will induce a phase on $|1\ra_{1c}|1\ra_{2c}$ state, and due to the finite time duration between each $\hat \Delta$ measurement, this phase can only be corrected up to a small deviation ranging between $[-\frac{\delta\theta}{2}, \frac{\delta \theta}{2}]$ where $\delta\theta = \frac{\pi}{2}\frac{\delta \tau}{T}$. This effect will on average give an error probability of $\frac{\kappa_1\gamma^2 (\delta \tau)^2 \pi^2}{384T}$ for $Z_1$, $Z_2$, and $Z_1Z_2$ type of errors. The second effect comes from both $\hat a_3$ and $\hat b_3$ happen in the same $\hat \Delta$ measurement interval. It can induce an effective $CZ_{1,2}(\pi\frac{T-t}{T})Z_3$ error with error rate $\kappa_1^2 \gamma^4 \delta \tau$. By averaging over time, this will give an error probability of $\frac{5}{8}\kappa_1^2 \gamma^4 \delta \tau T$ for $Z_3$ error and $\frac{1}{8}\kappa_1^2 \gamma^4 \delta \tau T$ for $Z_1Z_3$, $Z_2Z_3$ and $Z_1Z_2Z_3$ error.

In summary, the error probability for all types of errors can be listed as
% \begin{equation}
%     \begin{split}
%         & p_{Z_1} = p_{Z_2} = \kappa_1^2 \gamma^4 \delta \tau T + \frac{\pi^2}{256\kappa\gamma^6 T} + \frac{\kappa_1\gamma^2 (\delta \tau)^2 \pi^2}{384T},\\
%         & p_{Z_1Z_2} = \frac{\pi^2}{256\kappa\gamma^6 T} + \frac{\kappa_1\gamma^2 (\delta \tau)^2 \pi^2}{384T},\\
%         & p_{Z_3} = \frac{5}{8}\kappa_1^2 \gamma^4 \delta \tau T,\\
%         & p_{Z_1Z_3} = p_{Z_2Z_3} = p_{Z_1Z_2Z_3} = \frac{1}{8}\kappa_1^2 \gamma^4 \delta \tau T.
%     \end{split}
% \end{equation}

\begin{align*}
% \stepcounter{equation}\tag{\theequation}
\stepcounter{equation}
    & p_{Z_1} = p_{Z_2} = \kappa_1^2 \gamma^4 \delta \tau T + \frac{\pi^2}{256\kappa\gamma^6 T} + \frac{\kappa_1\gamma^2 (\delta \tau)^2 \pi^2}{384T},\\
    & p_{Z_1Z_2} = \frac{\pi^2}{256\kappa\gamma^6 T} + \frac{\kappa_1\gamma^2 (\delta \tau)^2 \pi^2}{384T},\\
    & p_{Z_3} = \frac{5}{8}\kappa_1^2 \gamma^4 \delta \tau T, \\
    & p_{Z_1Z_3} = p_{Z_2Z_3} = p_{Z_1Z_2Z_3} = \frac{1}{8}\kappa_1^2 \gamma^4 \delta \tau T.
    \tag{\theequation}
\end{align*}

To minimize the total error probability $p$ by using optimal choice of $T$, we can see that $p$ again scales as $p \sim O(\sqrt{\frac{\kappa_1}{\kappa}} \sqrt{\frac{\kappa_1 \delta\tau}{\gamma^2}}\sqrt{ 1 + C' \kappa\kappa_1\gamma^8 (\delta\tau)^2})$, which shares the same scaling as that in the case of the CNOT gate.

\nocite{*}

\bibliography{ref_paircat}

\end{document}